\documentclass[preprint,numerical,onecolumn,superscriptaddress]{revtex4-1}
\usepackage{bm}

\usepackage{amsmath}
\usepackage{amssymb}
\usepackage{graphicx}
\usepackage{comment}
\usepackage[colorlinks]{hyperref}

\begin{document}

\title{Derivation of the Hall and Extended Magnetohydrodynamics Brackets}
\author{Eric C. D'Avignon}
\email{cavell@physics.utexas.edu}
\author{Philip J. Morrison}
\email{morrison@physics.utexas.edu}
\affiliation{Department of Physics and Institute for Fusion Studies, The University of Texas at Austin, Austin, TX 78712, USA}
\author{Manasvi Lingam}
\email{mlingam@princeton.edu}
\affiliation{Department of Astrophysical Sciences, \\
Princeton University, Princeton, NJ 08544, USA}

\begin{abstract}
There are several plasma models intermediate in complexity between ideal magnetohydrodynamics (MHD) and two-fluid theory, with Hall and Extended MHD being two important examples.  In this paper we investigate several aspects of these theories, with the ultimate goal of deriving the noncanonical Poisson brackets used in their Hamiltonian formulations.  We present fully Lagrangian actions for each, as opposed to the fully Eulerian, or mixed Eulerian-Lagrangian, actions that have appeared previously.  As an important step in this process we exhibit each theory's two advected fluxes (in analogy to ideal MHD's advected magnetic flux), discovering also that with the correct choice of gauge they have corresponding Lie-dragged potentials resembling the electromagnetic vector potential, and associated conserved helicities.  Finally, using the Euler-Lagrange map, we show how to derive the noncanonical Eulerian brackets from canonical Lagrangian ones.
\end{abstract}

\maketitle

\section{Introduction}

Ideal magnetohydrodynamics (MHD), that reliable workhorse of plasma physics, has long been cast into noncanonical Hamiltonian form \cite{morrison80}.  So has the theory from which it is usually derived, the two-fluid model \cite{SK82}. There are many advantages to a Hamiltonian form: the discovery and classification of invariants; the development of numerical algorithms that automatically preserve such invariants; easily finding the equations of motion in curved coordinates; conducting equilibrium and stability analysis.  However, there are many theories intermediate in complexity between two-fluid theory and ideal MHD; Kimura and Morrison \cite{KM14} describe eleven of them.  Two are particularly important: Hall MHD, which accounts for the difference between the motion of the two species in a typical plasma, and L\"{u}st's Extended MHD \cite{L59}, which includes all terms of first order in the ratio $\mu$ of species masses in the derivation from two-fluid electron-ion theory.  Recently, Yoshida and Hameiri \cite{YH13} formulated a noncanonical Poisson bracket for Hall MHD, and shortly later Abdelhamid, Kawazura and Yoshida did so for Extended MHD \cite{abdelhamid15}; however, as often happens when working with Hamiltonian systems, they had to simply posit a bracket and prove it satisfied all the desired attributes, such as antisymmetry and the Jacobi identity.  We will show how to derive these brackets, starting from action principles for each theory.

These action principles have a long and distinguished history in fluids, originating with the work of Lagrange in the 18th century \cite{L18thcen}. The action principle formulation has also been employed in plasma physics since the second half of the 20th century, as evident from the works of \cite{Low58,Stu58,Su61,PH66,GK71,Dew70,Dew72,D74}. For ideal magnetohydrodynamics (MHD), the first action principle formulation was provided by Newcomb in \cite{Newcomb62}, shortly followed by other works in the same area \cite{Lund63,Cal63,Mer69}. For Extended MHD, an Eulerian action principle was proposed by \cite{IL99} which was subsequently generalized to a Eulerian-Lagrangian action by \cite{keramidas14}. For recent overviews of action principle formulations of plasma models, we refer the reader to \cite{mor05,mor09,MLA14}. The noncanonical Hamiltonian formulations for these models can be found in the works of \cite{morrison80,HK83,YH13,lingam15b,abdelhamid15,lingam15,LMM16}.

In principle there is an easy process to construct a noncanonical Poisson bracket, which goes as follows.  First, construct an action whose variations give the correct equations of motion in some coordinate system.  From this tangent-space action principle, derive a Hamiltonian function via a Legendre transform, and produce the corresponding phase-space action principle using the canonical momenta of the original action.  The Poisson bracket accompanying the phase-space action will be canonical.  Then simply change coordinates in order to produce the desired noncanonical bracket.  This procedure is, indeed, what we use, but there are many complications along the way.  

To begin with, the canonical bracket for fluid theories requires Lagrangian coordinates: those in which every fluid element is given a distinct label, and the equations of motion are expressed for a given labelled element, despite the fact that the element is changing position.  However, fully Lagrangian actions for Hall and Extended MHD have not yet been given.  The closest are the mixed Lagrangian-Eulerian actions of Ref. \cite{keramidas14}, ``Eulerian'' coordinates being ones that observe fluid quantities at a fixed point rather than following a given element.  In this paper we present fully Lagrangian actions.  Another complication arises because the Legendre transform fails to be invertible for either theory, and an expression for one of the velocities in terms of the phase-space variables must be inserted by hand.  Finally, the Euler-Lagrange map producing the noncanonical brackets requires prior knowledge of the generalized vorticities advected by the theories, so we must devote some time to their discovery and elucidation.

These generalized vorticities turn out to be crucial to the structure of every Hamiltonian MHD model.  There are $n$ advected vorticities for a theory with $n$ distinct charged species, $n$ being two in our case.  ``Advected'' in this case means that the flux elements defined by the vorticities are carried along with the fluid, their corresponding two-forms obeying a Lie-dragging equation.  For ideal MHD, both generalized vorticities collapse down to the same quantity, the magnetic field, which is advected by the fluid velocity.  For Hall MHD, one generalized vorticity is the magnetic field, whose fluxes are carried along with the electron velocity, and the other is the magnetic field plus kinetic vorticity $\nabla \times \mathbf{v}$, advected by the ion velocity \cite{MY98}.  For Extended MHD they turn out to be almost the same, but differing from the Hall MHD ones by terms of order $\mu$ in the curl of the current.  Both our actions and our derivations of noncanonical brackets would be impossible, but for the fact that we can eliminate the Eulerian magnetic field terms in both the action and the Euler-Lagrange map in favor of fully Lagrangian terms, an elimination wholly dependent on the existence of these advected fluxes.

Before moving on, we note a few ways in which our present work can be readily extended along the lines of past works that utilized these methods. One can incorporate finite Larmor radius (FLR) effects, such as the Braginskii gyroviscosity \cite{Brag65}.  FLR effects for reduced MHD \cite{MLA14} and generalized fluid models \cite{LMor14} were implemented via an action principle formulation, and evidently a similar treatment can be undertaken via our Extended MHD action principles.  Further extensions include stability analyses \cite{Mor82,HMRW85}, the systematic derivation of reduced Extended MHD models (with potential applications in collisionless reconnection), linear and nonlinear waves via a Lagrangian approach \cite{Whit74}, and MHD-like models (ideal, Hall, or Extended) for quasineutral plasmas with more than two charged species.

The paper is organized as follows.  Section \ref{subsec:HamSys} reviews the basic framework of Hamiltonian systems, while Section \ref{subsec:HamMHD} presents a specific example of that framework for ideal MHD, allowing comparisons with the related, but more complex constructions for Hall and Extended MHD.  We begin our new material by focusing on the simpler theory, Hall MHD, in Section \ref{sec:HallMHD}.  Section \ref{subsec:HallFlux} lays out the needed terminology and facts about Hall MHD, which are then used in Section \ref{subsec:HallActions} to construct both tangent-space and phase-space actions.  Section \ref{subsec:HallLieGauge} is an interesting digression in which we lay out a useful gauge, producing not only advected fluxes but corresponding advected one-forms.  Finally, we reach the goal which motivates our entire paper, the derivation in Section \ref{subsec:HallELMap} of the noncanonical bracket.  This derivation is carried out in more algebraic detail than might be necessary, in light of its unfamiliarity to many readers.  We then pivot to Extended MHD in Section \ref{sec:ExMHD}, starting with a derivation of its fluxes in Section \ref{subsec:ExMHDFlux}.  We give its actions in Section \ref{subsec:ExMHDAction}, and derive its noncanonical bracket in Section \ref{subsec:ExMHDELMap}.  This derivation takes more work than that in \ref{subsec:HallELMap}, but the procedure is identical, so this time we omit the details.  We conclude in Section \ref{sec:Conclusion}.

\section{Overview}
\label{sec:Overview}

\subsection{Hamiltonian Systems}
\label{subsec:HamSys}

As mentioned, all four models have been put into Hamiltonian form.  By that we mean that there is a functional $H$ and a bracket $\{f,g\}$ so that the time derivative of an arbitrary functional is given by
\begin{equation} \label{HamiltonsEquations}
\frac{d f}{d t} = \{f,H\} \; .
\end{equation}
The functionals are integral expressions of the field variables; for instance, the Hamiltonian functional is $H = \int \mathcal{H} \mathrm{d}^3 x$, where $\mathcal{H}$ is an energy density.  To separate out the time evolution of the field variables (like a momentum $m_i$), one can use a test functional such as
\begin{equation*}
\left. \frac{\partial m_i}{\partial t} \right|_{x_0} = \left\{\int m_i(x) \delta (x-x_0) \mathrm{d}^3 x,H\right\} \; .
\end{equation*}

The bracket can be expressed as
\begin{equation} \label{BracketExpansion}
\{f,g\} = \int \frac{\delta f}{\delta z_i}\mathcal{J}_{ij}(z) \frac{\delta g}{\delta z_j} \mathrm{d}^3 x \; ,
\end{equation}
where the $\delta / \delta z_i$ denotes functional differentiation and the components of $z$ represent the field variables.  The differential operators $\mathcal{J}_{ij}(z)$ must be chosen so that the bracket satisfies its usual properties (here $f$, $g$, and $h$ are functionals, and $\alpha$ and $\beta$ are real numbers):
\begin{align*}
& \{\alpha f + \beta g, h\} = \alpha \{f,h\} + \beta \{g,h\} & \\
& \{f,g\} = - \{g,f\} & \\
& \{\{f,g\},h\} + \{\{g,h\},f\} + \{\{h,f\},g\} = 0 \; . &
\end{align*}
Of these, only the last, the Jacobi identity, proves difficult to confirm.  Thankfully, the method of this paper provides a relatively easy way to confirm it.  In Lagrangian coordinates, where every fluid element is given a distinct label $a$ and the equations of motion are evaluated at fixed label (i.e. for a given fluid element), the bracket will be canonical:
\begin{equation*}
\{f,g\} = \int \left( \frac{\delta f}{\delta q^i}\frac{\delta G}{\delta \pi_i} -\frac{\delta g}{\delta q^i}\frac{\delta F}{\delta \pi_i} \right)
\mathrm{d}^3 a \; ,
\end{equation*}
where $q^i$ is the coordinate at fixed label $a$, and $\pi_i$ is its conjugate momentum, which can be obtained via an action.  This paper has such actions for each of its models.  

The Jacobi identity is fairly easy to prove for the canonical bracket, relying only on the commutation of functional derivatives, $\delta^2 f / \delta z^i \delta z^j = \delta^2 f / \delta z^j \delta z^i$.  However, the map converting Lagrangian to Eulerian coordinates, in which equations are expressed at fixed spatial coordinates, produces a noncanonical bracket; for example, even the straightforward definition
\begin{equation*}
m_i(x,t) = \int \pi_i(a,t) \delta(x - q(a,t)) \; \mathrm{d}^3 a
\end{equation*}
gives an Eulerian momentum dependent on both Lagrangian position and momentum.  The Jacobi identity can be directly proven for the Eulerian bracket, as was done in \cite{davignon15} for relativistic MHD and \cite{lingam15} for Hall MHD.  However, when such a bracket is produced from the canonical Lagrangian one, the Jacobi identity is assured, as it is invariant under coordinate changes and reductions.

\subsection{Hamiltonian MHD Models}
\label{subsec:HamMHD}

In this section we exhibit the already-discovered Hamiltonian forms for ordinary, Hall and Extended MHD.  It will be our goal in Secs. \ref{sec:HallMHD} and \ref{sec:ExMHD} to derive the latter two. These theories are all expressed here in terms of Eulerian variables, and Lagrangian equivalents will be postponed until our later discussion of their actions, where they occur more naturally.

In MHD the Eulerian field variables are density $\rho$, specific entropy $s$, fluid velocity $\mathbf{v}$ and magnetic field $\mathbf{B}$.  In the barotropic case, one can express $s$ as a function of $\rho$ and thereby eliminate it, but we consider the more general case.  One can also use as a supplementary variable the current density $(4 \pi / c) \mathbf{j} = \nabla \times \mathbf{B}$, using Amp\'{e}re's Law in the absence of displacement current.

Particle number and entropy are conserved and advected, respectively:
\begin{align*}
\frac{\partial \rho}{\partial t} + \left(\nabla \cdot \rho \mathbf{v} \right) = 0 \\
\frac{\partial s}{\partial t} + \mathbf{v}\cdot \nabla s = 0 \; .
\end{align*}
The fluid velocity obeys the following momentum equation:
\begin{equation} \label{EulersEquation}
\rho \left( \frac{\partial \mathbf{v}}{\partial t} + \mathbf{v}\cdot \nabla \mathbf{v} \right) = -\nabla p
+ \frac{\mathbf{j}\times\mathbf{B}}{c} \; ,
\end{equation}
while the magnetic field's evolution is determined by Ohm's Law for a perfect conductor:
\begin{equation*}
\mathbf{E} + \frac{\mathbf{v} \times \mathbf{B}}{c} = 0 \; ,
\end{equation*}
as can be seen by taking its curl and applying Faraday's Law:
\begin{equation*}
\frac{\partial \mathbf{B}}{\partial t} = \nabla \times \left(\frac{\mathbf{v} \times \mathbf{B}}{c} \right) \; .
\end{equation*}

In Hamiltonian MHD, while one can express the Hamiltonian and bracket in terms of $\mathbf{v}$ and $s$ (along with $\mathbf{B}$ and $\rho$), their derivation turns out to be simpler when using the momentum density $\mathbf{m} \equiv \rho \mathbf{v}$ and the entropy density $\sigma \equiv \rho s$.  In terms of these variables the Hamiltonian is the total energy
\begin{equation*}
H = \int \left(\frac{m^2}{2 \rho} + \rho U\left(\rho,\frac{\sigma}{\rho} \right) + \frac{B^2}{8 \pi}\right) \mathrm{d}^3 x
\end{equation*}
and the bracket is
\begin{align*}
\{f,g\}_{MHD} = & - \int \left(\rho \frac{\delta f}{\delta m_i}\frac{\partial}{\partial x^i} \left(\frac{\delta g}{\delta \rho}\right) 
- \rho \frac{\delta g}{\delta m_i}\frac{\partial}{\partial x^i} \left(\frac{\delta f}{\delta \rho}\right) \right) \\
& + \left(\sigma \frac{\delta f}{\delta m_i}\frac{\partial}{\partial x^i} \left(\frac{\delta g}{\delta \sigma}\right) 
- \sigma \frac{\delta g}{\delta m_i}\frac{\partial}{\partial x^i} \left(\frac{\delta f}{\delta \sigma}\right) \right) \\
& + \left(m_j \frac{\delta f}{\delta m_i}\frac{\partial}{\partial x^i} \left(\frac{\delta g}{\delta m_j}\right) 
- m_j \frac{\delta g}{\delta m_i}\frac{\partial}{\partial x^i} \left(\frac{\delta f}{\delta m_j}\right) \right) \\
& + \left(B^j \frac{\delta f}{\delta m_i}\frac{\partial}{\partial x^i} \left(\frac{\delta g}{\delta B^j}\right) 
- B^j \frac{\delta g}{\delta m_i}\frac{\partial}{\partial x^i} \left(\frac{\delta f}{\delta B^j}\right) \right) \\
& + \left(B^j \frac{\partial}{\partial x^j} \left(\frac{\delta f}{\delta B^i}\right) \frac{\delta g}{\delta m_i}
- B^j \frac{\partial}{\partial x^j} \left(\frac{\delta g}{\delta B^i}\right) \frac{\delta f}{\delta m_i} \right) \mathrm{d}^3 x \; .
\end{align*}
This bracket also constitutes the bulk of the Hall and Extended MHD brackets.  It was first given in Ref. \cite{morrison80}, and the sign convention used is from that paper.

Hall MHD differs from ordinary MHD in that the difference between ion and electron velocities is no longer neglected in Ohm's Law.  The derivation then modifies that Law to
\begin{equation} \label{HallOhmsLaw}
\mathbf{E} + \frac{\mathbf{v} \times \mathbf{B}}{c} = \frac{\mathbf{j} \times \mathbf{B}}{n e c} - \frac{\nabla p_e}{n e} \; ,
\end{equation}
where $n = \rho / m$ is the number density, $m$ is the particle mass (here equal to the ion mass) and $p_e$ is the electron pressure.  The entropy, continuity, and momentum equations are unchanged, as is the Hamiltonian.  The bracket, in turn, now has the additional term:
\begin{align}
\{f,g\} = & \; \{f,g\}_{MHD} + \{f,g\}_{Hall} \nonumber \\
= & \;  \{f,g\}_{MHD} - \int \frac{c}{ne} \mathbf{B} \cdot \left(\left(\nabla \times \frac{\delta f}{\delta \mathbf{B}}\right) \times 
\left(\nabla \times \frac{\delta g}{\delta \mathbf{B}}\right) \right) \mathrm{d}^3 x \label{HallBracket} \; .
\end{align}
This bracket was first described by \cite{YH13}, and it requires the assumption of a barotropic electron pressure.  Later we will see that this assumption gives us an advected magnetic flux, while barotropic ion pressure gives a second advected quantity.  These will be a necessary part of our construction of the bracket.

In Extended MHD, one additionally retains terms of first order in $\mu \equiv m_e / m_i$ during the derivation, producing a new momentum equation:
\begin{equation} \label{MomentumEquation}
nm \left(\frac{\partial \mathbf{v}}{\partial t} + \mathbf{v}\cdot\nabla \mathbf{v}\right) = - \nabla p + 
\frac{\mathbf{j}\times\mathbf{B}}{c} - \frac{m_e}{e^2}\mathbf{j}\cdot\nabla\left(\frac{\mathbf{j}}{n}\right)
\end{equation}
and a new version of Ohm's Law:
\begin{align} \label{OhmsLaw}
\mathbf{E} + \frac{\mathbf{v}\times\mathbf{B}}{c} = & \; \frac{m_e}{e^2 n} \left(\frac{\partial\mathbf{j}}{\partial t} 
+ \nabla\cdot\left(\mathbf{v}\mathbf{j} + \mathbf{j}\mathbf{v}\right) \right) \\
& \; - \frac{m_e}{e^3 n} \mathbf{j}\cdot\nabla\left(\frac{\mathbf{j}}{n}\right)
+ \frac{\mathbf{j}\times\mathbf{B}}{enc} - \frac{\nabla p_e}{en} \; . \nonumber
\end{align}
Here the term $\mathbf{v}\mathbf{j} + \mathbf{j}\mathbf{v}$ refers to the symmetric tensorial outer product.

The bracket and Hamiltonian, however, are more compactly expressed in terms of $\mathbf{B}_\star \equiv \mathbf{B} + (m_e c/e) \nabla \times \mathbf{v}$.  The Hamiltonian, which now includes a term for the electron kinetic energy, is
\begin{equation*}
H = \int \left(\frac{m^2}{2\rho} + \frac{n m_e}{2}\left(\frac{j}{ne}\right)^2 + \rho U(\rho,s) + \frac{B^2}{8\pi}\right)\mathrm{d}^3 x
= \int \left(\frac{m^2}{2 \rho} + \rho U(\rho,s) + \frac{1}{8\pi}\mathbf{B}\cdot\mathbf{B}_\star \right) \mathrm{d}^3 x
\end{equation*}
in light of the MHD Amp\'{e}re's Law $\nabla \times \mathbf{B} = (4\pi/c)\mathbf{j}$.  Its bracket, in turn, is
\begin{align}
\{f,g\} = & \; \{f,g\}_{MHD} + \{f,g\}_{Ext} \label{XMHDBracket} \\
= & \;  \{f,g\}_{MHD} - \int \frac{c}{ne} \left(\mathbf{B}_\star  - \frac{m_e c}{e} \nabla \times \frac{\mathbf{m}}{\rho}\right) \cdot 
\left(\left(\nabla \times \frac{\delta f}{\delta \mathbf{B}_\star}\right) \times 
\left(\nabla \times \frac{\delta g}{\delta \mathbf{B}_\star}\right) \right) \mathrm{d}^3 x \nonumber \; .
\end{align}
This bracket was first given in Ref. \cite{abdelhamid15}.  It also requires barotropic electron pressure $p_e = p_e(n)$.  The structural similarity between the Hall and Extended MHD Poisson brackets was investigated in Ref. \cite{lingam15,LMM16}, and will be elucidated further below.

\section{Hall MHD}
\label{sec:HallMHD}

\subsection{Flux conservation}
\label{subsec:HallFlux}

The essential difference between the various MHD models lies in their flux conservation laws, each one having a different version.  The archetypal flux conservation law is that of ordinary MHD, $\mathbf{B}\cdot \mathbf{d^2 q} = \mathbf{B_0} \cdot \mathbf{d^2 a}$ \cite{Newcomb62}.  Here the $a$ variables are coordinates in a label space $A$, whose continuous values identify fluid elements at $t = 0$ (this condition can be relaxed, as in \cite{AMP13}).  Meanwhile, the coordinates $q(a,t)$ describe the point to which a specific element flows; thus, $q(a,0) = a$.  In addition, $\mathbf{B} \equiv \mathbf{B}(q,t)$ while $\mathbf{B}_0 = \mathbf{B}_0(a) \equiv \mathbf{B}(q,0)$.  More explicitly, we write the flux conservation law as
\begin{equation} \label{OrdinaryFluxTwoForms}
\epsilon_{ijk}B^i (q,t) dq^j dq^k = \epsilon_{ijk} B^i_0 (a) da^j da^k \; .
\end{equation}
This expression can be manipulated into a transformation rule for the magnetic field:
\begin{equation} \label{OrdinaryFluxConservation}
B^i = \frac{B^j_0}{\mathcal{J}}\frac{\partial q^i}{\partial a^j}
\end{equation}
where $\mathcal{J} \equiv |\partial q / \partial a|$ is the Jacobian determinant of the invertible transformation from $a$ to $q$.
\begin{figure}[h]
\includegraphics{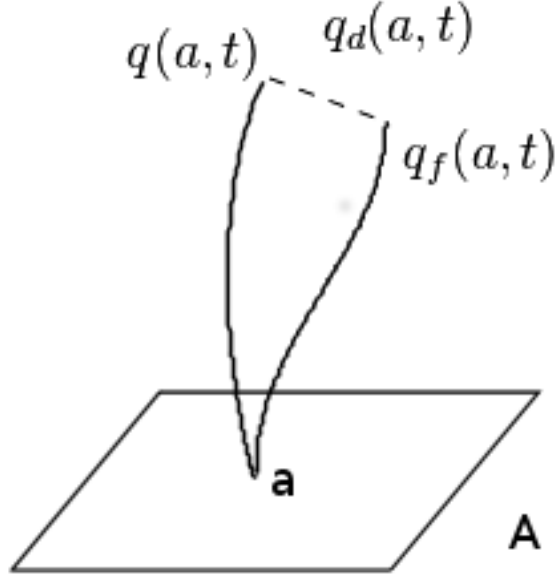}
\caption{Starting from label a, the fluid moves to $q(a,t)$, while the magnetic flux is dragged to $q_f(a,t)$.  Their difference is $q_d(a,t)$.}
\label{LabelSpaceFigure1}
\end{figure}

There are two distinct ways one can modify the flux conservation law \eqref{OrdinaryFluxTwoForms}.  First, one can advect a flux different from that of $\mathbf{B}$; with an appropriate choice of this flux, one then gets 2D inertial MHD \cite{lingam15b}.  Second, the same flux can be advected, but along a path distinct from that of the fluid.  This second approach gives Hall MHD.  Specifically, while the fluid itself flows from $a$ to a point $q(a,t)$, the flux element moves from $a$ to a distinct point $q_f(a,t)$, as illustrated by Figure \ref{LabelSpaceFigure1}.  Flux conservation is now
\begin{equation*}
\epsilon_{ijk}B^i dq_f^j dq_f^k = \epsilon_{ijk} B^i_0 \, da^j da^k \; ,
\end{equation*}
which gives rise to the transformation rule
\begin{equation} \label{FluxConservation}
B^i = \frac{B^j_0}{\mathcal{J}_f}\frac{\partial q_f^i}{\partial a^j} \; .
\end{equation}
The flux Jacobian $\mathcal{J}_f$ is also invertible, and can be written
\begin{equation*}
\mathcal{J}_f = \epsilon_{ijk}\epsilon^{lmn} \frac{\partial q_f^i}{\partial a^l} \frac{\partial q_f^j}{\partial a^m}
\frac{\partial q_f^k}{\partial a^n} \; ,
\end{equation*}
from which one can derive the expression $d \mathcal{J}_f / dt = \mathcal{J} \partial \dot{q}_f^i / \partial q_f^i$.

Taking a full time derivative of $\mathbf{B}(q_f, t)$ in equation \eqref{FluxConservation} gives
\begin{align*}
\frac{d B^i}{dt} = \frac{\partial B^i}{\partial t} + \dot{q}^j_f \frac{\partial B^i}{\partial q^j}
& =  \frac{B^j_0}{\mathcal{J}_f} \frac{\partial \dot{q}^i_f}{\partial a^j} -
\frac{B^j_0}{\mathcal{J}_f} \frac{\partial q^i_f}{\partial a^j} \frac{\partial \dot{q}^k_f}{\partial q^k_f} \\
& =  \frac{B^k_0}{\mathcal{J}_f}  \frac{\partial q^l_f}{\partial a^k} 
\frac{\partial a^m}{\partial q_f^l} \frac{\partial \dot{q}^i_f}{\partial a^m}
- \frac{B^j_0}{\mathcal{J}_f} \frac{\partial q^i_f}{\partial a^j} \frac{\partial \dot{q}^k_f}{\partial q^k_f} \\
& =  B^j \frac{\partial \dot{q}^i_f}{\partial q^j} - B^i \frac{\partial \dot{q}^k_f}{\partial q^k_f} \; .
\end{align*}
This equation shows that $\mathbf{B}$ is advected along $q_f$ as the vector dual to a 2-form, as desired.  Since $\mathbf{B}$ is divergenceless, we can add a term proportional to $(\nabla \cdot \mathbf{B})\mathbf{\dot{q}}_f$ and put the equation in the more familiar Faraday form
\begin{equation} \label{FaradaysLaw}
\frac{\partial \mathbf{B}}{\partial t} = \nabla_q \times \left(\mathbf{\dot{q}}_f \times \mathbf{B}\right) \; .
\end{equation}

\begin{figure}[h]
\includegraphics{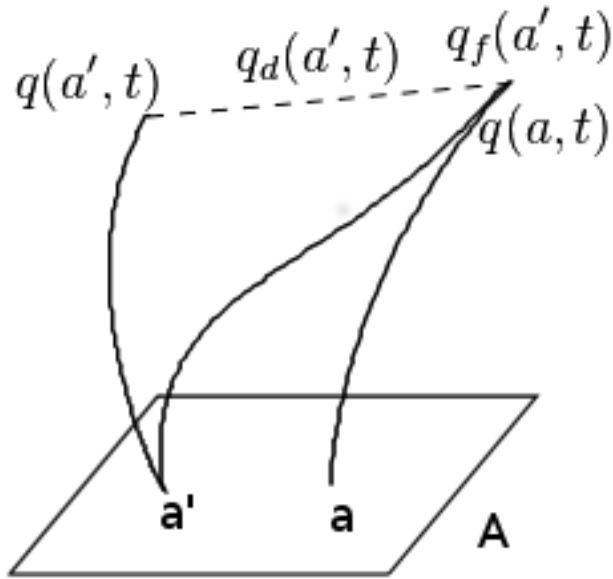}
\caption{The fluid moves to a point $q(a,t)$, where a flux has also flowed from a different initial point $a'$; thus $q_f(a',t) = q(a,t)$.}
\label{LabelSpaceFigure2}
\end{figure}

So far, so good.  However, complications arise when you look for the other equations of motion.  Some fluid attributes (density, specific entropy) are transported along the flow lines $q$, not $q_f$: mass conservation is described by $n(q,t) d^3 q = n_0(a) d^3 a$, and entropy conservation by $s(q,t) = s_0(a,t)$, recalling that no dissipative terms have been added.  As a result, the label corresponding to the magnetic field will differ from the label on the other quantities.  This situation is shown in Figure \ref{LabelSpaceFigure2}.
In this figure, the fluid element labelled by $a$ flows to $q(a,t)$, while a different label $a'$ shows the origin of the flux element that has been advected to $q(a,t) = q_f(a',t)$.  

For future use we will need two additional quantities: the point $q(a',t)$, to which the $a'$ element flows, and the difference $q_d(a',t)$ between $q_f(a',t)$ and $q(a',t)$.  All these quantities are related via
\begin{equation*}
q(a,t) = q_f(a',t) = q(a',t) + q_d(a',t) \; .
\end{equation*}
More relations are available, for example $a'(a,t) = q_f^{-1}(q(a,t),t)$.  In principle we could eliminate all but two of the quantities, but it is simpler to keep the extras around.  We also note that, for Hall MHD, $q$ corresponds to ion flow and $q_f$ to electron flow, so they might also have been written $q_i$ and $q_e$; however, in Extended MHD, we will use similar quantities, but they will now differ slightly from the electron and ion paths.  Thus we use a convention that will be appropriate for both models.  

We will also adopt the convention of using superscripts to show vectorial indices, and subscripts to show other attributes, like species identity or initial condition.  There are cases where the distinction between vectors and covectors (and thus raised and lowered indices) matters, such as when using curved coordinates, but it may easily be reinstated when needed.  We have, however, left the distinction intact in the expressions $q^i_{\; ,j}$ and $\delta^{ij}_{\; kl}$, where it improves readability.

\subsection{Lagrangian Actions}
\label{subsec:HallActions}

Every point corresponds to two labels.  In Hall MHD, using the convention described above, unprimed labels will denote ion quantities: for example, the number density is advected along the fluid lines, which in our approximation are the ion flow lines.  Meanwhile, primed labels will denote electron quantities, such as the magnetic flux density advected along electron flow lines.  In light of Fig. 2 above, the variable $q$ will appear as both $q(a,t)$ and $q(a',t)$, as will some quantities (namely the potentials $\phi$ and $\mathbf{A}$) dependent on them.  To simplify expressions we will write $q' \equiv q(a',t)$, $q'_d \equiv q_d(a',t)$, $(q')^i_{\; ,j} \equiv \partial (q')^i / \partial (a')^j$ and $(q_d')^i_{\; ,j} \equiv \partial (q_d')^i / \partial (a')^j$, with unprimed expressions such as $q$ denoting unprimed quantities like $q(a,t)$.  

If we treat primed and unprimed quantities separately, then the full Euler-Lagrange equations, using Lagrangian density $\mathcal{L}$, will be
\begin{align} \label{EulerLagrangeEquations}
\left[ \frac{d}{dt}\left(\frac{\partial \mathcal{L}}{\partial \dot{q}^i}\right)
+ \frac{d}{dt}\left(\frac{\partial \mathcal{L}}{\partial (\dot{q}')^i}\right)
+ \frac{\partial}{\partial a^j}\left(\frac{\partial \mathcal{L}}{\partial q^i_{\: ,j}}\right) \right.  & \\
\left. + \frac{\partial}{\partial (a')^j}\left(\frac{\partial \mathcal{L}}{\partial (q')^i_{\: ,j}}\right)
- \frac{\partial \mathcal{L}}{\partial q^i}
- \frac{\partial \mathcal{L}}{\partial (q')^i}  \right]_{a' = q_f^{-1}(q(a,t),t)} & = 0 \; . \nonumber
\end{align}
with a similar expression for $q_d$.  Many of the terms in the Euler-Lagrange equations are superfluous: only the first four terms will contribute in the $q$ variation, and only the second and fourth terms in the $q_d$ one.  These Euler-Lagrange equations can be obtained via Dirac delta function manipulations on a six-dimensional label space:
\begin{equation} \label{SixDAction}
S = \int \! \int \! \int \mathcal{L}(q,q_d,\dot{q},\dot{q}_d; q',q'_d,\dot{q}',\dot{q}'_d) \delta(a' - q^{-1}_f(q(a,t),t)) 
\; \mathrm{d}^3 a' \; \mathrm{d}^3 a \; \mathrm{d}t \, ;
\end{equation}
but for the most part, we will omit this consideration and work with \eqref{EulerLagrangeEquations}.  However, we emphasize one peculiarity of the action \eqref{SixDAction}: it only works if the delta function integral is performed after the variations.  If one does so before varying, collapsing back down to a single label space, the variational principle no longer gives the correct expressions.  This peculiarity is shared by the mixed Lagrangian-Eulerian approach of Ref. \cite{keramidas14}.

If it were written in terms of ion and electron velocities $q_i$ and $q_e$, the Lagrangian density would be standard:
\begin{align} \label{ElectronIonLagrangian}
\mathcal{L} = & \; \frac12 m_i n_0 \dot{q}^2_i + \frac{e n_0}{c} \mathbf{\dot{q}}_i \cdot \mathbf{A}(q,t) - e n_0 \phi (q,t) - 
n_0 U_i \left(\frac{n_0}{\mathcal{J}}, s_{(0)i}\right) \\ 
& + \frac12 m_e n_0 (\dot{q}'_e)^2 - \frac{e n_0}{c} \mathbf{\dot{q}}'_e \cdot \mathbf{A}(q' + q'_d,t) + e n_0 \phi (q' + q'_d,t) \nonumber
- n_0 U_e \left(\frac{n_0}{\mathcal{J}_f}, s_{(0)e}\right) \; .
\end{align}
In Hall MHD, we treat electron velocity as being different from ion velocity (unlike in regular MHD), but nonetheless neglect terms of order $m_e / m_i$.  The variables used will be center-of-mass velocity $\dot{q}$, and the drift velocity of electrons relative to ions, $\dot{q}_d$.  In terms of ion and electron velocities we have
\begin{equation*}
\mathbf{\dot{q}} = \frac{m_i \mathbf{\dot{q}}_i + m_e \mathbf{\dot{q}}_e}{m_i + m_e} \qquad
\mathbf{\dot{q}}_d = \mathbf{\dot{q}}_e - \mathbf{\dot{q}}_i \; .
\end{equation*}
Inverting these equations and neglecting terms of the order of the mass ratio, we have
\begin{align}
\mathbf{\dot{q}}_i = \mathbf{\dot{q}} - \frac{m_e}{m_i + m_e} \mathbf{\dot{q}}_d \approx & \; \mathbf{\dot{q}} 
\label{CoordinateChange} \\
\mathbf{\dot{q}}_e = \mathbf{\dot{q}} + \frac{m_i}{m_i + m_e} \mathbf{\dot{q}}_d \approx & \; \mathbf{\dot{q}} + \mathbf{\dot{q}}_d \; .
\nonumber
\end{align}
Thus, rewriting \eqref{ElectronIonLagrangian}, setting $m \equiv m_i + m_e \approx m_i$, and noting the distinction between primed and unprimed labels, the Lagrangian density becomes
\begin{align}
\mathcal{L} = & \; \frac12 m n_0 \dot{q}^2 + \frac{e n_0}{c} \left[ \mathbf{\dot{q}} \cdot \mathbf{A}(q,t) - \mathbf{\dot{q}'} \cdot \mathbf{A}(q' + q'_d, t)  - \mathbf{\dot{q}'}_d \cdot \mathbf{A}(q' + q'_d, t) \right] \nonumber \\
&  - e n_0 \left[\phi(q,t) - \phi(q' + q'_d,t) \right] 
 - n_{0} \left[ U_i  \left( \frac{n_{0}}{\mathcal{J}}, s_{(0)i} \right) + U_e \left(\frac{n_0}{\mathcal{J}_f}, s_{(0)e} \right) \right] \; .
\label{TangentSpaceAction}
\end{align}

In the $q$ equation of motion, the terms arising from $\phi(q,t) - \phi(q' + q'_d,t)$ cancel, plus most of the terms coming from $\mathbf{\dot{q}} \cdot \mathbf{A}(q,t) - \mathbf{\dot{q}'} \cdot \mathbf{A}(q' + q'_d, t)$, due to the $q = q' + q'_d$ evaluation.  The only surviving term comes from the advective parts of $d \mathbf{A} / dt$, which are different for the two terms.  An additional term arises from the $q'$ partial derivative on $q'_d \cdot \mathbf{A}(q' + q'_d)$.  Setting $p_e = n^2 \partial U_e / \partial n$, $p_i = n^2 \partial U_i / \partial n$, and $p = p_e + p_i$, we have, for the $q$ equation of motion,
\begin{equation*}
\left[ m n_0 \ddot{q}^i - \frac{e n_0}{c} \left(\dot{q}'_d \right)^j \partial^j A^{i}(q' + q'_d,t) 
+ \frac{e n_0}{c} \left(\dot{q}'_d \right)^j \partial^i A^{j}(q' + q'_d,t) + \mathcal{J} \partial^i p \right]_{a' = q_f^{-1}(q(a,t),t)} = 0
\end{equation*}
which can, by multiplying with $1 / \mathcal{J}$ and using $\mathbf{j} = - e n \dot{\mathbf{q}}_d$, be simplified to
\begin{equation} \label{MomentumEquationLag}
\rho \mathbf{\ddot{q}} = - \nabla_q \, p + \frac1c \mathbf{j} \times \mathbf{B} \; ,
\end{equation}
which is the Lagrangian equivalent of \eqref{EulersEquation}.

In the $q_d$ equation of motion, the three final terms come from the full derivative $d \mathbf{A}(q' + q_d', t) / dt$, and the pressure term comes from the $q'_d$ dependence of $\mathcal{J}_f$:
\begin{align*}
\frac{e n_0}{c} \left( \left(\dot{q}'\right)^j \partial^i A^{j} (q' + q'_d, t) +  \left(\dot{q}'_d \right)^j \partial^i A^{j} (q' + q'_d, t) \right)
- e n_0 \partial^i \phi (q' + q'_d, t) \\
+ \mathcal{J} \partial^i p_e - \frac{e n_0}{c} \left( (\dot{q}')^j \partial^j A^{i} (q' + q'_d, t) + (\dot{q}'_d)^j \partial^j A^{i} (q' + q'_d, t) \right) - 
\frac{\partial A^i}{\partial t} = 0
\end{align*}
with the whole thing evaluated at $q(a,t) = q(a',t) + q_d(a',t)$ as usual.  Reordering and simplifying, one finds
\begin{equation}
\mathbf{E} + \frac{\mathbf{\dot{q}} \times \mathbf{B}}{c} = - \frac{\dot{\mathbf{q}_d} \times \mathbf{B}}{c} - \frac{\nabla p_e}{n e} \; ,
\end{equation}
which is Ohm's Law \eqref{HallOhmsLaw} for Hall MHD.  Finally, the canonical momenta are
\begin{equation} \label{HallCanonicalMomenta}
\mathbf{\pi} = \frac{\delta L}{\delta \mathbf{\dot{q}}} = m n_0 \dot{\mathbf{q}} \qquad \qquad 
\mathbf{\pi}_d = \frac{\delta L}{\delta \mathbf{\dot{q}}_d} = -\frac{e n_0}{c} \mathbf{A} \; ,
\end{equation}
with the Lagrangian function $L$ defined by $S = \int L\; \mathrm{d}t$.  The expression for $\pi_d$ will allow us to convert the thus-far omitted field term $B^2 / 8 \pi$ into a term expressed by fluid quantities, once we switch to a phase-space action.

It turns out to be easy to translate the mixed Eulerian-Lagrangian terms of Ref. \cite{keramidas14} into the appropriate terms of \eqref{HallOhmsLaw}, when one minds the difference between primed and unprimed labels.  We translate that paper's $Q$ and $D$ into our variables using $q = Q$ and $q_d = -D/en_0$.  Then its mixed ion terms in the Lagrangian are
\begin{align*}
& \int \! \int e n_0 \left( \frac1{c} \dot{\mathbf{q}}\cdot\mathbf{A}(x,t) - \phi(x,t)\right) \delta(x-q(a,t)) \; d^3 x \; d^3 a \\
= & \int \! \int \! \int e n_0 \left(\frac1{c} \dot{\mathbf{q}}\cdot\mathbf{A}(x,t) - \phi(x,t)\right) \delta(x-q(a,t))
\; \delta (a' - q_f^{-1}(q(a,t),t)) \; d^3 x \; d^3 a \; d^3 a' \\
= & \int \! \int e n_0 \left( \frac1{c} \dot{\mathbf{q}}\cdot\mathbf{A}(q,t) - \phi(q,t)\right)
\; \delta (a' - q_f^{-1}(q(a,t),t)) \; d^3 a \; d^3 a'
\end{align*}
and its electron terms are
\begin{align*}
& \int \! \int e n_0 \left( \frac1{c} \left(\dot{\mathbf{q}}+ \dot{\mathbf{q}}_d\right) \cdot\mathbf{A}(x,t) - \phi(x,t)\right) 
\delta(x-q(a',t) - q_d(a',t)) \; d^3 x \; d^3 a' \\
= & \int \! \int \! \int e n_0 \left(\frac1{c} \left(\dot{\mathbf{q}}+ \dot{\mathbf{q}}_d\right) \cdot\mathbf{A}(x,t) - \phi(x,t)\right) 
\delta(x-q(a',t) - q_d(a',t)) \\
& \times \; \delta (a' - q_f^{-1}(q(a,t),t)) \; d^3 x \; d^3 a' \; d^3 a \\
= & \int \! \int e n_0 \left( \frac1{c}  \left(\dot{\mathbf{q}}+ \dot{\mathbf{q}}_d\right) \cdot\mathbf{A}(q' + q'_d,t) - \phi(q'+q'_d,t)\right)
\; \delta (a' - q_f^{-1}(q(a,t),t)) \; d^3 a \; d^3 a' \; .
\end{align*}

However, the action in Ref. \cite{keramidas14} also contains a fully Eulerian term
\begin{equation*}
\int \frac{-1}{8\pi} \left| \nabla \times \mathbf{A}(x,t) \right|^2 \; d^3 x \; ,
\end{equation*}
which is used to produce $-(4\pi/c) \mathbf{\dot{q}}_d = \nabla \times \mathbf{B}$, a missing piece in our fully Lagrangian tangent space action.  We also cannot perform the usual Legendre transform, because we have no expression $\dot{q}_d(q,q_d, \pi, \pi_d)$.  Fortunately, we can solve these problems by switching to a phase space action and invoking \eqref{FluxConservation}.  The four variations of this action give all the needed equations.  The needed action density is
\begin{align}
\mathbf{\pi} \cdot \mathbf{\dot{q}} + \mathbf{\pi}_d \cdot \mathbf{\dot{q}}_d - \frac{1}{2 m n_0} \pi^2
+ \frac{e}{mc} \left(\mathbf{\pi}\cdot \mathbf{A}(q,t) - \mathbf{\pi}\cdot \mathbf{A}(q' + q'_d,t) \right) \nonumber \\
- \frac{1}{8 \pi \mathcal{J}_f} \left(\frac{c}{n_0 e} \right)^2
\left(\nabla_a \times \mathbf{\pi}_{(0)d} \right)^i \left(\nabla_a \times \mathbf{\pi}_{(0)d} \right)^j
\left(\frac{\partial (q')^k}{\partial (a')^i} + \frac{\partial (q'_d)^k}{\partial (a')^i} \right)
\left(\frac{\partial (q')^k}{\partial (a')^j} + \frac{\partial (q'_d)^k}{\partial (a')^j} \right) \nonumber \\
+ n_0 e \left(\phi(q' + q'_d, t) - \phi(q,t) \right) - 
n_0 \left[ U_i  \left( \frac{n_{0}}{\mathcal{J}}, s_{(0)i} \right) + U_e \left(\frac{n_0}{\mathcal{J}_f}, s_{(0)e} \right) \right] \; . \label{PhaseSpaceDensity}
\end{align}
The middle term, note, is simply $B^2 / 8 \pi$.  We have expanded it using \eqref{FluxConservation} to express the magnetic field in terms of its initial value, and then applying \eqref{HallCanonicalMomenta} to express this initial value as the curl of that of a canonical momentum.

There are four phase space variations; as when using \eqref{EulerLagrangeEquations}, one sets $q = q' + q'_d$ after taking variations.  Thus the $\pi$ variation gives
\begin{equation} \label{PiVariation}
\mathbf{\dot{q}} = \frac{\mathbf{\pi}}{m n_0} \; .
\end{equation}
The $\pi_d$ variation involves an integration by parts on the middle term of the density \eqref{PhaseSpaceDensity}, giving
\begin{equation} \label{PiDVariation}
\mathbf{\dot{q}}_d = - \frac{c}{4 \pi n_0 e} \nabla \times \mathbf{B} \; ,
\end{equation}
i.e. $(4 \pi / c) \mathbf{j} = \nabla \times \mathbf{B}$, the missing piece of our earlier tangent space action.  Note that the $B^2/8\pi$ term requires varying $\pi_{(0)d}$, which is the value of $\pi_d$ on the boundary at $t=0$.  This is permitted since the action principle only requires $\delta q = \delta q_d = 0$ on the boundary in order to perform an integration by parts, while the momenta are free to vary at $t = 0$.

Once again, most of the terms vanish in the $q$ variation.  The $\partial q / \partial a$ terms in the middle term of \eqref{PhaseSpaceDensity} give two factors of $\partial^j (B^i B^j /2)$, and the $\mathcal{J}_f$ in the same term gives a factor of $\partial^i (B^2 / 2)$.  The remaining terms proceed similarly as in our tangent space calculation.  The overall result is
\begin{equation} \label{QVariation}
- \mathbf{\dot{\pi}}^i - \frac{C^{kj}}{4\pi} \frac{\partial}{\partial a^k} \left[B^i B^j - \frac{B^2}{2} \delta^{ij} \right] + \mathcal{J} \partial^i p = 0 \; ,
\end{equation}
where $C^{jk}$ is the cofactor matrix to $\partial q^i / \partial a^j$.   Given $\mathbf{j} = (c/\pi) \nabla \times \mathbf{B}$ and the $\epsilon$-$\epsilon$ identity, \eqref{QVariation} is the same as \eqref{MomentumEquationLag}.  Finally, the $q_d$ variation gives
\begin{equation} \label{QDVariation}
- \dot{\pi}_d^i - \frac{e}{mc} \pi^j \partial^i A_j + \frac{C^{kj}}{4\pi} \frac{\partial}{\partial a^k} \left[B^i B^j - \frac{B^2}{2} \delta^{ij} \right]
+ n_0 e\partial^i \phi + \mathcal{J}_f \partial^i p_e = 0 \; .
\end{equation}
Considering that $\pi_d = -(e n_0 / c) \mathbf{A}(q,t)$, and $\dot{\pi}_d$ will thus have two terms, this equation is identical to \eqref{HallOhmsLaw}.

However, the action \eqref{PhaseSpaceDensity} is still slightly unsatisfactory, because we use the quantities $\mathbf{A}$ and $\phi$, which are not fully determined by \eqref{FluxConservation}: namely, their gauge freedom remains.  We did use the relation (\ref{HallCanonicalMomenta}), viz. $\pi_d = -(e n_0 / c) \mathbf{A}(q,t)$, from the tangent space action \eqref{TangentSpaceAction} to construct the phase space action \eqref{PhaseSpaceDensity}; however, \eqref{PhaseSpaceDensity} does not produce this relation, and neither action gives us the evolution of $\phi$.  Sec. \ref{subsec:HallLieGauge} develops a gauge condition which resolves this problem in an elegant manner.

\subsection{The Lie gauge and advection of the vector potential}
\label{subsec:HallLieGauge}

Look at the Hall MHD Ohm's Law \eqref{HallOhmsLaw} in Eulerian coordinates:
\begin{equation*} 
\mathbf{E} + \frac{\mathbf{v} \times \mathbf{B}}{c} = \frac{\mathbf{j} \times \mathbf{B}}{n e c} - \frac{\nabla p_e}{n e} \; .
\end{equation*}
Using $\mathbf{E} = -\nabla \phi - (1/c) \partial \mathbf{A} / \partial t$ and reordering, it becomes
\begin{equation} \label{VectorEquation}
\frac{\partial \mathbf{A}}{\partial t} = \left(\mathbf{v} - \frac{\mathbf{j}}{ne}\right) \times \mathbf{B}
+ \frac{c}{ne} \nabla p_e - c \nabla \phi \quad 
\end{equation}
and, for a barotropic plasma in which $p_e = p_e(n)$, taking the curl renders it into the form
\begin{equation*}
\frac{\partial \mathbf{B}}{\partial t} = \nabla \times \left(\mathbf{v}_f \times \mathbf{B}\right)
\end{equation*}
with $\mathbf{v}_f = \mathbf{v} - \mathbf{j}/ne$.  This equation is in the form of \eqref{FaradaysLaw}, showing that the components of $\mathbf{B}$ are dual to those of a two-form which is Lie-dragged by $\mathbf{v}_f$.  

It would be even more convenient for $\mathbf{A}$ to be the components of a Lie-dragged one-form, with $\mathbf{B}$ dual to the components of its exterior derivative.  Because the last two terms of \eqref{VectorEquation} are curl-free, they can be expressed as a gradient: $\nabla \phi' = \nabla \phi - (1/ne) \nabla p_e$.  We then use the gauge freedom in $\phi$ to set
\begin{equation} \label{LieGauge}
\nabla \phi' = \nabla \left( \frac{\mathbf{v}_f \cdot \mathbf{A}}{c} \right)
\end{equation}
which we call the Lie gauge, due to the fact that it will produce a Lie-dragging equation.  With this gauge equation \eqref{VectorEquation} becomes
\begin{align*}
\frac{\partial \mathbf{A}}{\partial t} = & \; \mathbf{v}_f \times \left(\nabla \times \mathbf{A}\right) - 
\nabla \left( \mathbf{v}_f \cdot \mathbf{A} \right) \\
= & \; v_{(f)j} \partial_i A_j - v_{(f)j} \partial_j A_i - (\partial_i v_{(f)j}) A_j - v_{(f)j} (\partial_i A_j) \\
= & \; -v_{(f)j} \partial_j A_i - (\partial_i v_{(f)j}) A_j
\end{align*}
or
\begin{equation*}
\frac{\partial \mathbf{A}}{\partial t} + \pounds_{\mathbf{v}_f} \mathbf{A} = 0 \; ,
\end{equation*}
so that the vector potential is now a Lie-dragged one-form, as desired.

In fact, there exists an entire family of gauges that result in a Lie-dragged one-form.  Suppose $\mathbf{A}$ is one such member (like that already provided), and $\psi$ is a Lie-dragged zero-form, so that $\psi(q_f,t) = \psi_0(a)$ and
\begin{equation} \label{PsiAdvected}
\frac{\partial  \psi }{\partial t}+ \mathbf{v}_f \cdot \nabla \psi= \frac{\partial  \psi }{\partial t}+ \pounds_{\mathbf{v}_f}  \psi=0\,,
\end{equation}
Let $\mathbf{A} = \mathbf{A}' + \nabla \psi$.  Then, starting from 
\begin{equation*}
\frac{\partial \mathbf{A}}{\partial t} = \; \mathbf{v}_f \times \left(\nabla \times \mathbf{A}\right) - 
\nabla \left( \mathbf{v}_f \cdot \mathbf{A} \right) \; ,
\end{equation*}
we have
\begin{equation*}
\frac{\partial \mathbf{A}'}{\partial t} + \frac{\partial \nabla \psi}{\partial t} = \; \mathbf{v}_f \times \left(\nabla \times \mathbf{A}'\right) - 
\nabla \left( \mathbf{v}_f \cdot \mathbf{A}' \right) - \nabla\left(\mathbf{v}_f \cdot \left(\nabla \psi \right)\right) \; .
\end{equation*}
Collecting the $\psi$ terms inside an overall gradient operator and applying \eqref{PsiAdvected} eliminates all of them, showing that $\mathbf{A}'$ is also an advected one-form.

Lie-dragging of $\mathbf{A}$ as a one-form implies that $A^i \: dq_f^i = A^i_{0} \: da^i$, thus
\begin{equation} \label{VectorPotentialTransformation}
A^j \frac{\partial q^j}{\partial a^i} = A^i_{0} \enspace \Rightarrow \enspace A^i = A^j_{0} \frac{\partial a^j}{\partial q_f^i} = 
\frac{A^j_{0} C^{ji}_{f}}{\mathcal{J}_f} \; ,
\end{equation}
where $C_f^{ji}$ is the cofactor matrix of the coordinate transformation $\partial q_f^j / \partial a^i$.  Because of the relation $\mathbf{A} = - (c / e n_0) {\pi}_d $, the canonical momentum also transforms as a one-form:
\begin{equation} \label{PiTransformationLaw}
\pi^i_{d} = \pi^j_{0,d} \frac{C_f^{ji}}{\mathcal{J}_f} \; ,
\end{equation}
Using the Lie gauge \eqref{LieGauge}, one can eliminate $\phi$ from the phase space action \eqref{PhaseSpaceDensity}, and using \eqref{PiTransformationLaw} one can also eliminate $\mathbf{A}(q' + q'_d,t)$ in favor of its initial value $\mathbf{A}'_0$ at $t=0$.

However, the other appearance of the vector potential is $\mathbf{A}(q,t)$ in \eqref{TangentSpaceAction} is written in terms of ``ion quantities'' (i.e. unprimed variables), whereas \eqref{VectorPotentialTransformation} expresses it using solely the ``electron quantity'' (i.e. primed variable) $q'_f \equiv q' + q'_d$.  Thus, we've only solved half the problem: we've expressed $A(q'+q'_d,t)$ in terms of $A_0(a',t)$ and eliminated $\phi(q'+q'_d,t)$ with a gauge condition, but we're still left with $A(q,t)$ and $\phi(q,t)$.  Thankfully, there is a general result showing that, in a system of $n$ charged fluid species with barotropic equations of state, there are $n$ conserved helicities \cite{mahajan02} and $n$ Lie-dragged two-forms \cite{MLin15}.  In MHD, this duality is of no concern, because the two collapse and give rise to a single magnetic helicity $\int \mathbf{A}\cdot\mathbf{B} \; d^3x$; however, in more general models they remain distinct.  We can use the other helicity to eliminate the last two extraneous variables.

Hall MHD has the following variable as its second Lie-dragged two-form:
\begin{equation} \label{HallIonVorticity}
\mathbf{\mathcal{B}} = \mathbf{B} + \frac{cm}{e} \nabla \times \mathbf{v} \; .
\end{equation}
Its advection is straightforward to prove, using Eulerian variables.  Taking the time derivative, and remembering our assumption of barotropic pressures,
\begin{align*}
\frac{\partial \mathbf{\mathcal{B}}}{\partial t} = & \; \frac{\partial \mathbf{B}}{\partial t} + \frac{cm}{e} \nabla \times \frac{\partial \mathbf{v}}{\partial t} \\
=& \; -\pounds_{\mathbf{v}_f} \mathbf{B} + \frac{cm}{e} \nabla \times \left[\mathbf{v} \times \left(\nabla \times \mathbf{v}\right)
- \nabla \left(\frac12 v^2\right) - \frac{\nabla p}{mn} + \frac{\mathbf{j}\times\mathbf{B}}{mcn} \right] \\
= & \; \nabla \times \left[ \mathbf{v} \times \left(\mathbf{B} + \frac{cm}{e} \nabla \times \mathbf{v} \right) \right]
= \nabla \times \left(\mathbf{v} \times \mathbf{\mathcal{B}} \right) = -\pounds_{\mathbf{v}} \mathbf{\mathcal{B}} \; .
\end{align*}

Since $\mathbf{\mathcal{B}}$ is divergenceless, it can be expressed as the curl of a vector $\mathbf{\mathcal{A}}$.  A fully general expression for such a vector is
\begin{equation} \label{HallIonPotential}
\mathbf{\mathcal{A}} = \mathbf{A} + \frac{cm}{e} \mathbf{v} + \nabla \psi \; .
\end{equation}
Just as $\phi$ was for $\mathbf{A}$, $\psi$ can be chosen to make $\mathbf{\mathcal{A}}$ a Lie-dragged one-form, expressed as
\begin{equation} \label{OtherPotentialTransformation}
\mathcal{A}_i = \frac{\mathcal{A}_{(0)j} C^j_{\; i}}{\mathcal{J}} \; .
\end{equation}
For we have
\begin{align*}
\frac{\partial \mathbf{\mathcal{A}}}{\partial t} = & \; \frac{\partial \mathbf{A}}{\partial t} + \frac{cm}{e} \frac{\partial \mathbf{v}}{\partial t}
+ \nabla\left(\frac{\partial \psi}{\partial t}\right) \nonumber \\
= & \; \left(\mathbf{v}_f \times \mathbf{B} + \frac{c}{ne} \nabla p_e - c \nabla \phi \right) \nonumber \\
& + \frac{cm}{e} \left( \mathbf{v} \times \left(\nabla \times \mathbf{v}\right) - \nabla \left(\frac{1}{2}v^2\right) - \frac{\nabla p}{mn}
+ \frac{\mathbf{j}\times\mathbf{B}}{mnc} \right) + \nabla \left(\frac{\partial \psi}{\partial t}\right) \nonumber \\
= & \; \mathbf{v} \times \left(\nabla \times \mathbf{\mathcal{A}} \right) +
\left[ \nabla \left( \frac{\partial \psi}{\partial t} - c \phi - \left(\frac12 \frac{cm}{e} v^2\right) \right)
+\frac{c}{ne}\left(\nabla p_e - \nabla p\right) \right] \; .
\nonumber
\end{align*}
Following the reasoning that motivated the Lie gauge \eqref{LieGauge}, we note that the expression in square brackets is the equivalent of $- c \nabla \phi'$ from before.  We can thus get a Lie-dragged one-form $\mathbf{\mathcal{A}}$ by setting
\begin{equation} \label{UsefulForSegregation}
\nabla \left[\frac{1}{c} \frac{\partial \psi}{\partial t} - \phi - \left(\frac12 \frac{m}{e} v^2\right) \right]
+ \frac{1}{ne} \nabla p_e - \frac{1}{ne} \nabla p = 
- \nabla \left(\frac{\mathbf{v} \cdot \mathbf{\mathcal{A}}}{c}\right) \; .
\end{equation}
Using the Lie gauge \eqref{LieGauge}, this equation simplifies to
\begin{equation*}
\frac{\partial \psi}{\partial t} + \frac{c}{e} \left(\frac12 m v^2 + \frac{\mathbf{j}\cdot\mathbf{A}}{ne} - \frac{p}{n} \right) = 0 \; ,
\end{equation*}
solved by the action-like quantity
\begin{equation} \label{PseudoAction}
\psi = -\frac{c}{e} \int_{t_0}^t \left( \frac12 m v^2 + \frac{\mathbf{j}\cdot\mathbf{A}}{ne} - \frac{p}{n} \right) d^3 x
\end{equation}
Just as before, you can add a Lie-dragged zero-form to $\psi$ and still have $\mathbf{\mathcal{A}}$ Lie-dragged.

However, tantalizing though the action-like expression \eqref{PseudoAction} is, we needed an action expressed entirely in terms of ion quantities and electron quantities, while the above one is mixed due to $\mathbf{j}\cdot \mathbf{A}$ being an electron quantity.  Thus we go back to line \eqref{UsefulForSegregation}, which can be simplified slightly to 
\begin{equation*}
\frac{\partial \psi}{\partial t} - c \phi - \left(\frac12 \frac{cm}{e} v^2\right) - \frac{cp_i}{ne} = 
- \mathbf{v} \cdot \mathbf{\mathcal{A}} \; ,
\end{equation*}
where $p_i$ is the ion pressure.  Except for $\phi$, which is still an electron quantity, and $\psi$, which is mixed, these are all ion quantities.  Thus we can create the following quantity:
\begin{equation*}
\Upsilon \equiv \phi - \frac1c \frac{\partial \psi}{\partial t} = - \left(\frac12 \frac{m}{e} v^2\right) + 
\frac{\mathbf{v} \cdot \mathbf{\mathcal{A}}}{c} - \frac{p_i}{n} \; ,
\end{equation*}
which is an ion quantity because all the terms on the right hand side are.  With the four quantities $\mathbf{A}$, $\mathcal{A}$, $\phi$ and $\Upsilon$, obeying transformation rules like \eqref{VectorPotentialTransformation} and subject to the Lie gauge, the potential terms in the action \eqref{PhaseSpaceDensity} can be expressed entirely in terms of their initial conditions, solving the problem mentioned at the end of Sec. \ref{subsec:HallActions}.

The new ion variables $\mathcal{A}$ and $\Upsilon$ introduced in this section deserve a bit more attention.  By writing the expression
\begin{equation*}
- \frac{\partial \mathcal{A}}{\partial t} - \nabla \Upsilon = \mathbf{E} - \frac{m}{e} \frac{\partial \mathbf{v}}{\partial t} \equiv \mathcal{E}
\end{equation*}
we can say, loosely, that $\Upsilon$ is to $\mathbf{\mathcal{A}}$ what $\phi$ is to $\mathbf{A}$.  The parallel is further reinforced by the equivalent of Faraday's Law,
\begin{equation*}
\nabla \times \mathcal{E} = \frac{1}{c} \frac{\partial \mathcal{B}}{\partial t}
\end{equation*}
and an easily derived expression
\begin{equation} \label{ExtraOhmsLaw}
\mathcal{E} + \frac{\mathbf{v}_f \times \mathcal{B}}{c} = -\frac{1}{nec} \mathbf{j}\times\mathcal{B} + \frac{\nabla p_i}{ne}
+ \frac{m}{e} \nabla \left(\frac12 v^2\right)
\end{equation}
reminiscent of Ohm's Law, but with an extra gradient term.  Were all the time-dependent terms to be removed from $\mathcal{E}$, \eqref{ExtraOhmsLaw} would be a generalization of Bernoulli's Law.

We conclude this section with two observations.  First, we can combine \eqref{VectorPotentialTransformation}, \eqref{FluxConservation}, and $d^3 q' = \mathcal{J}_f \; d^3 a'$ to show that
\begin{equation*}
\mathbf{A}\cdot\mathbf{B} \: d^3 q' = \mathbf{A}_0 \cdot \mathbf{B}_0 \: d^3 a' \; ,
\end{equation*}
which is a nicely compact proof of the conservation of magnetic helicity.  The similar expressions derivable for the ion quantities can similarly be combined to read
\begin{equation*}
\mathbf{\mathcal{A}}\cdot\mathbf{\mathcal{B}} \: d^3 q = \mathbf{\mathcal{A}}_0 \cdot \mathbf{\mathcal{B}}_0 \: d^3 a \; .
\end{equation*}
Second, everything that has been said in this section applies to ideal MHD as well, which simply requires one to use $\mathbf{v}$ instead of $\mathbf{v}_f$ and to remember that there is only one distinct helicity.

\subsection{Euler-Lagrange Map and the derivation of the Eulerian bracket}
\label{subsec:HallELMap}

Our phase-space action principle may equivalently be expressed as the set of Hamilton's equations $\partial f/ \partial t = \{f,H\}$, for arbitrary functionals $f$ of the phase-space variables.  The bracket in this case is the canonical one:
\begin{equation} \label{CanonicalBracket}
\{f,g\} = \int \left(\frac{\delta f}{\delta q^i}\frac{\delta g}{\delta \pi^i}
- \frac{\delta g}{\delta q^i}\frac{\delta f}{\delta \pi^i}
+ \frac{\delta f}{\delta q_d^i}\frac{\delta g}{\delta \pi_d^i}
- \frac{\delta g}{\delta q_d^i}\frac{\delta f}{\delta \pi_d^i}\right) d^3 a \; .
\end{equation}
In this section we will show how to convert this bracket into the noncanonical bracket \eqref{HallBracket}.

The Eulerian quantities $\rho$, $\sigma$, and $m^i$ are defined via standard Euler-Lagrange maps:
\begin{align}
\rho(x,t) = \int \rho(a,t) \: \delta (x & - q(a,t)) \: d^3 q = \int \rho_0(a) \: \delta(x - q(a,t)) \: d^3 a \nonumber \\
\sigma(x,t) = & \int \rho_0(a) s_0(a) \: \delta(x-q(a,t)) \: d^3 a \label{OtherEulerLagrangeMaps} \\
m^i(x,t) = & \int \pi^i(a,t) \: \delta(x-q(a,t)) \: d^3 a \nonumber \; .
\end{align}
The variable $\sigma$ is superfluous for barotropic Hall MHD, but it is included here for the sake of generality.  When we induce variations later on, the quantities $\rho$ and $\sigma$ will only have $\delta q$ variations (from the delta functions), while $\mathbf{m}$ will have a $\delta q$ and $\delta \pi$ variation.  The odd one is the magnetic Euler-Lagrange map:
\begin{align} 
B^i(x,t) & = \int B_0^j (a') \frac{\partial q_f^i}{\partial (a')^j} \delta(x - q_f(a',t)) \: d^3 a' \nonumber \\
& = \int B_0^j (a') \left(\frac{\partial q^i}{\partial (a')^j} + \frac{\partial q_d^i}{\partial (a')^j} \right) 
\delta(x - q(a',t) - q_d(a',t)) \: d^3 a' \; .
\label{MagneticEulerLagrange}
\end{align}
This will have $q$ and $q_d$ dependence via $q_f$, and $\pi_d$ dependence via $\pi_d = - (e n_0 / c) \mathbf{A}$.

We can now show how the Eulerian variables change under variations in the Lagrangian phase-space ones, using \eqref{OtherEulerLagrangeMaps} and \eqref{MagneticEulerLagrange}:
\begin{align}
\delta \rho = & - \int \rho_0(a) \: \delta'_i (x - q(a,t)) \: \delta q^i \: d^3 a \nonumber \\
\delta \sigma = & - \int \sigma_0(a) \: \delta'_i (x-q(a,t)) \: \delta q^i \: d^3 a \label{Variations} \\
\delta m^i = & - \int \pi^i \: \delta'_j (x-q(a,t)) \: \delta q^j + \delta(x-q) \: \delta \pi^i \: d^3 a \nonumber \\
\delta B^i = & \int - B_0^j (a') \left(\frac{\partial q^i}{\partial (a')^j} + \frac{\partial q_d^i}{\partial (a')^j} \right)
\delta'_k(x-q(a',t)-q_d(a',t)) \left(\delta q^k + \delta q^k_d \right) \nonumber \\
& + B_0^j \delta'_k \left(x - q(a',t) - q_d(a',t)\right) 
\left(\frac{\partial q^k}{\partial (a')^j} + \frac{\partial q_d^k}{\partial (a')^j} \right) \left( \delta q^i + \delta q_d^i \right) \nonumber \\
& + \frac{\partial B_0^j}{\partial \pi_d^k} \delta \pi_d^k
\left(\frac{\partial q^i}{\partial (a')^j} + \frac{\partial q_d^i}{\partial (a')^j} \right) \delta(x - q' - q'_d) \: d^3 a' \; . \nonumber
\end{align}
Note that the addition of $q_d$ and $\pi_d$, which do not appear in regular MHD, nonetheless do not require us to add any new Eulerian variables.  They do, however, add a new term in the variation $\delta B^i$ that does not appear in ideal MHD, because now $\mathbf{B}$ has a $\pi_d$ dependence via $\mathbf{B} = \nabla \times \mathbf{A}$.

The variation induced by an arbitrary function $f$, in both Lagrangian and Eulerian variables, is
\begin{align}
\delta f = & \int \frac{\delta f}{\delta \rho} \delta \rho + \frac{\delta f}{\delta \sigma} \delta \sigma
+ \frac{\delta f}{\delta m^i} \delta m^i + \frac{\delta f}{\delta B^i} \delta B^i \: d^3 x \nonumber \\ \label{ComparisonOfVariations}
= & \int \frac{\delta f}{\delta q^i} \delta q^i + \frac{\delta f}{\delta \pi^i} \delta \pi^i
+ \frac{\delta f}{\delta q_d^i} \delta q_d^i + \frac{\delta f}{\delta \pi_d^i} \delta \pi_d^i \: d^3 a \; .
\end{align}
Substituting the various \eqref{Variations}, except for the one term involving $\delta \pi_d$ (which will require more careful attention), into the left side of \eqref{ComparisonOfVariations} gives the expression
\begin{align*}
& - \int \!\! \int \left[
\left(\frac{\delta f}{\delta \rho}\rho_0(a) + \frac{\delta f}{\delta \sigma}\sigma_0(a) + \frac{\delta f}{\delta m^i}\pi^i\right)
\delta'_j \left(x - q(a,t)\right) \delta q^j \right. \\
& + \frac{\delta f}{\delta B^i} 
\left(B^j_0 (a)\frac{\partial q^i_f}{\partial a^j} \delta q^k - B^j_0 (a) \frac{\partial q^k_f}{\partial a^j} \delta q^i\right)
\delta'_k \left(x - q(a,t) - q_d(a,t) \right) \\
& + \frac{\delta f}{\delta B^i}
\left(B^j_0 (a) \frac{\partial q^i_f}{\partial a^j} \delta q_d^k - B^j_0 (a) \frac{\partial q^k_f}{\partial a^j} \delta q_d^i\right)
\delta'_k \left(x - q(a,t) - q_d(a,t) \right) \\
& + \left. \left(\frac{\delta f}{\delta m^i} \delta \left(x - q(a,t) \right)\right) \delta \pi^i \right]  d^3 x \: d^3 a \; .
\end{align*}
In this expression, the disappearance of $a'$ is rather startling, but it is still there implicitly via the delta functions, for at a fixed $x$ they will pick out values of $a$ for the magnetic terms distinct from those of the other terms.

Meanwhile, the term that we omitted is, by using $\mathbf{B}_0 = \nabla_a \times \mathbf{A}_0$, given by
\begin{align*}
& - \int \!\! \int \frac{\delta f}{\delta B^i} \epsilon^{jkl} \frac{\partial}{\partial a^k} \left( \frac{c}{n_0 e} \delta \pi^l_{d,0} \right) \frac{\partial q^i_f}{\partial a^j} \:
\delta(x-q-q_d) \: d^3 a \: d^3 x\\
=  & \enspace \int \!\! \int \frac{c}{n_0 e} \frac{\delta f}{\delta B^i} \epsilon^{jkl} \;  \delta {\pi}^l_{d,0} \frac{\partial q^i_f}{\partial a^j}
\frac{\partial q^m_f}{\partial a^k} \delta'_{m} (x-q-q_d) \: d^3 a \: d^3 x \; .
\end{align*}
Here the $\partial^2 q_f / \partial a \partial a$ term in the integration by parts vanishes because it is a symmetric object contracted with an antisymmetric one, and the second factor of $\partial q_f / \partial a$ appears because we want the delta-function derivative to give a derivative with respect to $q$ (and thus $x$).  These factors may be eliminated in the following manner:
\begin{align*}
\epsilon^{jkl} \frac{\partial q^i_f}{\partial a^j} \frac{\partial q^m_f}{\partial a^k} 
= & \; \frac12 \epsilon^{jkl} \left(\frac{\partial q^i_f}{\partial a^j} \frac{\partial q^m_f}{\partial a^k} 
- \frac{\partial q^i_f}{\partial a^k} \frac{\partial q^m_f}{\partial a^j} \right) \\
= & \; \frac12 \epsilon^{jkl} \frac{\partial q^a_f}{\partial a^j} \frac{\partial q^b_f}{\partial a^k} \delta^{im}_{\: ab}
= \; \frac12 \epsilon^{jkl} \frac{\partial q^a_f}{\partial a^j} \frac{\partial q^b_f}{\partial a^k} \epsilon^{nim} \epsilon^{nab} \\
= & \; \frac12 C^{ln} \epsilon^{nim} \; .
\end{align*}
Thus, using \eqref{PiTransformationLaw}, that portion of the $\delta f$ variation becomes
\begin{equation*}
\int \!\! \int \frac{c}{2 n_0 e} \frac{\delta f}{\delta B^i} \mathcal{J}_f \: \delta \pi^j_{d} \: \epsilon^{jik}
\delta'_{k} (x - q - q_d) \: d^3 a \: d^3 x \; .
\end{equation*}
Comparison of the expanded Eulerian $\delta f$ with the right side of \eqref{ComparisonOfVariations} then gives expressions for the Lagrangian functional derivatives in terms of the Eulerian ones:
\begin{align*}
\frac{\delta f}{\delta \pi^i} = & \int \frac{\delta f}{\delta m^i} \delta\left( x-q(a,t) \right) d^3 x = 
\left. \frac{\delta f}{\delta m^j} \right|_{x = q(a,t)} \\
\frac{\delta f}{\delta q^i} = & - \int \left(\frac{\delta f}{\delta \rho} \rho_0 + \frac{\delta f}{\delta \sigma} \sigma_0
+ \frac{\delta f}{\delta m^i} \pi \right) \delta'_i \left(x - q \right) \\
& + \frac{\delta f}{\delta B^j} B^k_0 \frac{\partial q^j_f}{\partial a^k} \delta'_i \left(x - q - q_d \right)
- \frac{\delta f}{\delta B^i} B^k_0 \frac{\partial q^j_f}{\partial a^k} \delta'_j \left(x - q - q_d \right) d^3 x \\
 & =  \int \left[ \rho_0 \frac{\partial}{\partial x^i} \left(\frac{\delta f}{\delta \rho}\right) 
 + \sigma_0 \frac{\partial}{\partial x^i} \left(\frac{\delta f}{\delta \sigma}\right) 
 + \pi^j \frac{\partial}{\partial x^i} \left(\frac{\delta f}{\delta m^j}\right) \right] \delta(x-q)  \\
 & + \mathcal{J}_f \left[ B^j \frac{\partial}{\partial x^i} \left(\frac{\delta f}{\delta B^j}\right) 
  - B^j \frac{\partial}{\partial x^j} \left(\frac{\delta f}{\delta B^i}\right) \right] \delta(x-q-q_d) \: d^3 x \\
\frac{\delta f}{\delta q_d^i} = & - \int 
\frac{\delta f}{\delta B^j} B^k_0 \frac{\partial q^j_f}{\partial a^k} \delta'_i \left(x - q - q_d \right)
- \frac{\delta f}{\delta B^i} B^k_0 \frac{\partial q^j_f}{\partial a^k} \delta'_j \left(x - q - q_d \right) d^3 x \\
= & \int \mathcal{J}_f \left[ B^j \frac{\partial}{\partial x^i} \left(\frac{\delta f}{\delta B^j}\right) 
  - B^j \frac{\partial}{\partial x^j} \left(\frac{\delta f}{\delta B^i}\right) \right] \delta(x-q-q_d) \: d^3 x \\
\frac{\delta f}{\delta \pi^i_d} = & \int \frac{\delta f}{\delta B^j} \frac{c}{2 n_0 e} \mathcal{J}_f \: \epsilon^{ijk} \delta'_{k} (x - q - q_d) \: d^3 x \\
= & \; \frac{c}{2 n e} \int \left(\nabla \times \frac{\delta f}{\delta \mathbf{B}} \right) _i \:
\delta (x - q - q_d) \: d^3 x = - \frac{c}{2 n e} \left. \left(\nabla \times \frac{\delta f}{\delta \mathbf{B}} \right) _i \;
\right|_{x = q(a,t) + q_d(a,t)}
\end{align*}

Finally, we can insert these functional derivatives into the canonical Lagrangian bracket \eqref{CanonicalBracket}.  Evaluating the delta function introduces a factor of $\mathcal{J}^{-1}$ or $\mathcal{J}^{-1}_f$, eliminates the $d^3 a$ integral and converts the remaining Lagrangian quantities into Eulerian ones:
\begin{align}
\{f,g\} = & \int \left(\frac{\delta f}{\delta q^i}\frac{\delta g}{\delta \pi^i} -
\frac{\delta g}{\delta q^i}\frac{\delta f}{\delta \pi^i} \right) +
\left(\frac{\delta f}{\delta q_d^i}\frac{\delta g}{\delta \pi_d^i} -
\frac{\delta g}{\delta q_d^i}\frac{\delta f}{\delta \pi_d^i} \right) \: d^3 a \nonumber \\
= & - \int \left(\rho \frac{\delta f}{\delta m^i}\frac{\partial}{\partial x^i} \left(\frac{\delta g}{\delta \rho}\right) 
- \rho \frac{\delta g}{\delta m^i}\frac{\partial}{\partial x^i} \left(\frac{\delta f}{\delta \rho}\right) \right) \nonumber \\
& + \left(\sigma \frac{\delta f}{\delta m^i}\frac{\partial}{\partial x^i} \left(\frac{\delta g}{\delta \sigma}\right) 
- \sigma \frac{\delta g}{\delta m^i}\frac{\partial}{\partial x^i} \left(\frac{\delta f}{\delta \sigma}\right) \right) \nonumber \\
& + \left(m_j \frac{\delta f}{\delta m^i}\frac{\partial}{\partial x^i} \left(\frac{\delta g}{\delta m_j}\right) 
- m_j \frac{\delta g}{\delta m^i}\frac{\partial}{\partial x^i} \left(\frac{\delta f}{\delta m_j}\right) \right) \nonumber \\
& + \left(B^j \frac{\delta f}{\delta m^i}\frac{\partial}{\partial x^i} \left(\frac{\delta g}{\delta B^j}\right) 
- B^j \frac{\delta g}{\delta m^i}\frac{\partial}{\partial x^i} \left(\frac{\delta f}{\delta B^j}\right) \right) \label{BigBracket}\\
& + \left(B^j \frac{\partial}{\partial x^j} \left(\frac{\delta f}{\delta B^i}\right) \frac{\delta g}{\delta m^i}
- B^j \frac{\partial}{\partial x^j} \left(\frac{\delta g}{\delta B^i}\right) \frac{\delta f}{\delta m^i} \right) \nonumber\\
& + \frac{c}{2 n e}  \left[B^j \left(\nabla \times \frac{\delta f}{\delta \mathbf{B}}\right)^i \frac{\partial}{\partial x^i} \left(\frac{\delta g}{\delta B^j}\right) 
- B^j \left(\nabla \times \frac{\delta g}{\delta \mathbf{B}}\right)^i \frac{\partial}{\partial x^i} \left(\frac{\delta f}{\delta B^j}\right)  \right.  \nonumber \\
& + \left. B^j \frac{\partial}{\partial x^j} \left(\frac{\delta f}{\delta B^i}\right) \left(\nabla \times \frac{\delta g}{\delta \mathbf{B}}\right)^i
- B^j \frac{\partial}{\partial x^j} \left(\frac{\delta g}{\delta B^i}\right) \left(\nabla \times \frac{\delta f}{\delta \mathbf{B}}\right)^i  \right] d^3 x \nonumber \\
\equiv & \quad \{f,g\}_{MHD} + \{f,g\}_{Hall} \nonumber \; .
\end{align}
Here the $\{f,g\}_{Hall}$ terms are those in the square bracket, and the remaining $\{f,g\}_{MHD}$ terms are familiar from ordinary MHD.

The Hall portion of the bracket can be greatly simplified.  Take the two terms involving the curl of $\delta f / \delta \mathbf{B}$.  They become
\begin{align*}
& \frac{c}{2 n e} \left[B^j \left(\nabla \times \frac{\delta f}{\delta \mathbf{B}}\right)^i \; \delta^{kl}_{\;ij} \:
\frac{\partial}{\partial x^k}\left(\frac{\delta g}{\delta B^l}\right)\right] \\
= \; & \frac{c}{2 n e} \left[B^j \left(\nabla \times \frac{\delta f}{\delta \mathbf{B}}\right)^i \; \epsilon_{mij}\epsilon^{mkl}
\frac{\partial}{\partial x^k}\left(\frac{\delta g}{\delta B^l}\right)\right] \\
= \; & - \frac{c}{2 n e} \mathbf{B} \cdot \left[\left(\nabla \times \frac{\delta f}{\delta \mathbf{B}}\right) \times
\left(\nabla \times \frac{\delta g}{\delta \mathbf{B}}\right)\right] \; .
\end{align*}
The other two terms give an identical expressions; together, they eliminate the factor of $1/2$ and reproduce the Hall MHD bracket \eqref{HallBracket}.

Before we move on to produce results for Extended MHD, we should pause a moment to discuss our peculiar method of introducing a phase-space constraint.  The simplest phase-space action is a finite-dimensional particle one extremizing
\begin{equation*}
S = \int_{t_0}^{t_f} \sum_{i} \mathbf{\dot{q}}_{(i)} \cdot \mathbf{p}_{(i)} - H(q,p) \; dt
\end{equation*}
with fixed endpoints $t_0$ and $t_f$.  When doing the $q$ variations, an integration by parts must be performed, so $\delta q = 0$ on the endpoints of the action integral; however, when varying $p$, no integration by parts is required, so the momenta can vary on the endpoints.  In our Lagrangian density \eqref{PhaseSpaceDensity}, the $q$, $\pi$, and $q_d$ variations occur as normal.  However, $\pi_d$ has been expressed entirely in terms of its initial value $\pi_{(0)d}$.  Thus, when doing the $\pi_d$ variation, one \emph{only} varies at the endpoints (here the initial surface $t=0$), with the variation at $t>0$ determined, ultimately, by \eqref{FluxConservation} via \eqref{HallCanonicalMomenta}.  The same substitution for $\pi_d$ in terms of its initial value occurs in the magnetic Euler-Lagrange map \eqref{MagneticEulerLagrange}, making it crucial for the derivation of the Hall MHD bracket.  We consider the successful derivation of the bracket to be a sign of this constraint's validity.  However, viewing it as a specific instance of a more general method (hopefully with applications elsewhere in Hamiltonian physics), it is clear that we have not established the full conditions under which this method may be applied.  We hope to do so in future work.  

\section{Extended MHD}
\label{sec:ExMHD}

\subsection{Advected quantities}
\label{subsec:ExMHDFlux}

In the course of writing an action for Extended MHD, we will need to write the field portions in terms of an advected two-form, which will be a vorticity-like quantity.  Unfortunately, this time around the magnetic field $\mathbf{B}$ is not such a quantity.  We will thus begin by showing how to derive a pair of vorticity equations in Extended MHD.  Written in a standard fashion, the two central equations for Extended MHD are the momentum equation \eqref{MomentumEquation} and the generalized Ohm's Law \eqref{OhmsLaw}, rounded off with Ampere's Law and the continuity equation, plus the isentropy equation if needed.

Our goal will be to use these equations to derive a pair of vorticity equations
\begin{equation} \label{VorticityEquations}
\frac{\partial \mathbf{B_{\pm}}}{\partial t} = \nabla \times \left(\mathbf{v}_\pm \times \mathbf{B}_\pm \right) \; .
\end{equation}
Previous experience with ideal and Hall MHD suggests that there should be such equations, along with the result that a theory of $n$ charged fluids will have $n$ such vorticities. Moreover, it was shown in Ref. \cite{lingam15} that they do exist, by exploiting a map between the Poisson brackets for Hall and Extended MHD.  However, our purpose in this paper is to derive those very brackets, so we should not rely on knowledge drawn from those brackets.  Thus we will show how to derive the equations \eqref{VorticityEquations} directly.

In Ref. \cite{keramidas14}, the form \eqref{OhmsLaw} of the generalized Ohm's Law is derived from the equivalent expression
\begin{align*}
\mathbf{E} + \frac{\mathbf{v}\times\mathbf{B}}{c} = & \; \frac{m_e}{e^2 n} \left(\frac{\partial\mathbf{j}}{\partial t} 
+ \mathbf{j}\cdot\nabla\mathbf{v} - \mathbf{j}\cdot\nabla\left(\frac{\mathbf{j}}{ne}\right)
+ \left(\nabla\cdot\mathbf{v}\right)\mathbf{j}\right) \\
& \; + \frac{\mathbf{j}\times\mathbf{B}}{enc} - \frac{\nabla p_e}{en} + 
\frac{m_e}{e^2}\mathbf{v}\cdot\nabla\left(\frac{\mathbf{j}}{n}\right) +
\frac{m_e}{n^2 e^2} \mathbf{j} \left(\mathbf{v}\cdot\nabla\right)n
\end{align*}
by combining the last two terms and adding a term proportional to $\nabla\cdot\mathbf{j}$ (which is zero).  Instead of doing that, we combine the first term, the last term, and the term proportional to $(\nabla\cdot\mathbf{v}) \mathbf{j}$ into $(m_e / e^2) (\partial / \partial t) (\mathbf{j} / n)$ by using the continuity equation.  We can then replace all occurrences of $\mathbf{j}$ with $\mathbf{u} \equiv \mathbf{j} / (ne)$, which has units of velocity, producing
\begin{align*}
\mathbf{E} + \frac{\mathbf{v}\times\mathbf{B}}{c} = & \; \frac{m_e}{e} \left( \frac{\partial \mathbf{u}}{\partial t}
+ \mathbf{u}\cdot\nabla\mathbf{v} + \mathbf{v}\cdot\nabla\mathbf{u}
- \mathbf{u}\cdot\nabla\mathbf{u}\right) \\
& \; + \frac{\mathbf{u}\times\mathbf{B}}{c} - \frac{\nabla p_e}{n e} \; .
\end{align*}
We next apply the $\nabla(\mathbf{A}\cdot\mathbf{B})$ identity and switch to the new field variable $\mathbf{B}_\star \equiv \mathbf{B} + (m_e c / e) \nabla \times \mathbf{u}$ to produce
\begin{align} \nonumber
\mathbf{E} + \frac{\mathbf{v}\times\mathbf{B}_\star}{c} = & \; \frac{m_e}{e} \left[ \frac{\partial \mathbf{u}}{\partial t}
+ \nabla\left(\mathbf{u}\cdot\mathbf{v} - \frac12 u^2\right) - \mathbf{u}\times\left(\nabla\times\mathbf{v}\right)\right] \\
& \; + \frac{\mathbf{u}\times\mathbf{B}_\star}{c} - \frac{\nabla p_e}{n e} \; .
\label{OhmsLawU}
\end{align}
Performing similar operations on the momentum equation \eqref{MomentumEquation} produces
\begin{align} \label{MomentumEquationU}
\frac{\partial\mathbf{v}}{\partial t} - \mathbf{v}\times\left(\nabla\times\mathbf{v}\right) =
- \frac{\nabla p}{nm} - \nabla \left(\frac12 \frac{m_e}{m} u^2 + \frac12 v^2 \right) + \frac{e}{mc} \mathbf{u}\times\mathbf{B}_\star \; .
\end{align}
Equations \eqref{OhmsLawU} and \eqref{MomentumEquationU}, being slightly more compact than \eqref{MomentumEquation} and \eqref{OhmsLaw}, suggest that $\mathbf{u}$ and $\mathbf{B}_\star$ are indeed more natural variables than $\mathbf{j}$ and $\mathbf{B}$.

The various gradients in \eqref{OhmsLawU} and \eqref{MomentumEquationU} are crying out for us to take a curl, so we will.  Using Faraday's Law and collecting the time derivative terms in one place, \eqref{OhmsLawU} becomes
\begin{equation} \label{BEquation}
\frac{\partial \mathbf{B}_\star}{\partial t} = \nabla \times \left[\frac{m_e c}{e} \mathbf{u}\times\left(\nabla\times\mathbf{v}\right)
+ \mathbf{v}\times\mathbf{B}_\star - \mathbf{u}\times\mathbf{B}_\star \right] 
- \frac{c}{ne^2} \left(\nabla p_e \times \nabla n\right)
\end{equation}
while \eqref{MomentumEquationU} becomes
\begin{equation} \label{VEquation}
\frac{\partial}{\partial t}\left(\nabla \times \mathbf{v}\right) = \nabla \times \left[ \mathbf{v}\times\left(\nabla\times\mathbf{v}\right)
+ \frac{e}{mc} \mathbf{u}\times\mathbf{B}_\star \right] + \frac1{mn^2} \left(\nabla p \times \nabla n\right) \; .
\end{equation}
Going forward we assume barotropic equations of state for both the electrons and ions, because then $\nabla p_s \propto \nabla n$ and the pressure terms drop.

In \eqref{BEquation} and \eqref{VEquation}, shorn of their pressure terms, we have all the ingredients we need to make two equations of the form \eqref{VorticityEquations}.  We assume that the quantities appearing in the vorticity equations are linear combinations of those that have appeared in deriving \eqref{BEquation} and \eqref{VEquation}:
\begin{align} \nonumber
\mathbf{B}_\pm = & \; \delta_\pm \mathbf{B}_\star + \beta_\pm \frac{mc}{e} \left(\nabla \times \mathbf{v}\right) \\
\mathbf{v}_\pm = & \; \gamma_\pm \mathbf{v} + \alpha_\pm \mathbf{u}
\label{LinearCombination}
\end{align}
The coefficients are all unitless.  Expanding \eqref{VorticityEquations}, we have
\begin{align} \nonumber
\frac{\partial \mathbf{B}_\pm}{\partial t} = \nabla \times & \left[ 
\; \gamma_\pm \delta_\pm \left(\mathbf{v}\times\mathbf{B}_\star\right)
+ \alpha_\pm \delta_\pm \left(\mathbf{u}\times\mathbf{B}_\star\right) \right. \\
& \; \left. + \beta_\pm \gamma_\pm \frac{mc}{e} \left(\mathbf{v}\times \left(\nabla \times \mathbf{v}\right) \right)
+ \beta_\pm \alpha_\pm \frac{mc}{e} \left(\mathbf{u}\times\left(\nabla \times \mathbf{v}\right)\right) \right] \; .
\label{VorticityExpansion}
\end{align}
On the other hand, we can also use linear combinations of \eqref{BEquation} and \eqref{VEquation} to express $\partial \mathbf{B}_\pm / \partial t$.  Equating the resulting coefficients with what we find in \eqref{VorticityExpansion}, we have the system of equations
\begin{align*}
\delta_\pm = \delta_\pm \gamma_\pm \qquad \qquad & \beta_\pm - \delta_\pm = \alpha_\pm \delta_\pm \\
\beta_\pm = \beta_\pm \gamma_\pm \qquad \qquad & \frac{m_e}{m} \delta_\pm = \beta_\pm \alpha_\pm \; .
\end{align*}
One equation is redundant, which should be no surprise since $\mathbf{B}_\pm$ is only established up to an overall scale.  We use this extra freedom to set $\delta_\pm = 1$.  Then the solutions are $\gamma_\pm = 1$, and
\begin{equation} \label{Coefficients}
\alpha_\pm = \frac12 \left(-1 \pm \sqrt{1 + 4 \mu}\right) \qquad \qquad \beta_\pm = \frac{1}{2} \left(1 \pm \sqrt{1 + 4 \mu}\right) = - \alpha_\mp \; ,
\end{equation}
where $\mu = m_e / m_i$ is the electron-ion mass ratio.  This confirms what was found in Ref. \cite{lingam15}.  We can also invert \eqref{LinearCombination} to get
\begin{align} \label{BStarFromBPM}
\mathbf{B}_\star = & \; \frac{\beta_+ \mathbf{B}_- - \beta_- \mathbf{B}_+}{\beta_+ - \beta_-}
& \; \nabla \times \mathbf{v} = &\; \frac{e}{mc} \frac{\mathbf{B}_+ - \mathbf{B}_-}{\beta_+ - \beta_-} \\
\mathbf{v} = & \; \frac{\alpha_+ \mathbf{v}_- - \alpha_- \mathbf{v}_+}{\alpha_+ - \alpha_-}
& \; \mathbf{u} = &\; \frac{\mathbf{v}_+ - \mathbf{v}_-}{\alpha_+ - \alpha_-} \; . \nonumber
\end{align}

Because the $\mathbf{B}_\pm$ are divergenceless, they can each be written as the curl of a vector potential $\mathbf{A}_\pm$.  Accounting for the gauge freedom, these potentials are
\begin{equation} \label{ExtendedPotentials}
\mathbf{A}_\pm = \mathbf{A} + \frac{m_e c}{e} \mathbf{u} + \beta_\pm \frac{m c}{e} \mathbf{v} + \nabla \psi_\pm \; .
\end{equation}
Thus, taking appropriate linear combinations of \eqref{OhmsLawU} and \eqref{MomentumEquationU}, we find
\begin{align*}
\frac{1}{c} \frac{\partial \mathbf{A}_\pm}{\partial t} = & \;
\frac{\mathbf{v}_\pm \times \mathbf{B}_\pm}{c} 
- \frac{m}{e}\nabla \left(\frac12 \beta_\pm v^2 + \frac{m_e}{m} \mathbf{u}\cdot\mathbf{v} - \frac{m_e}{m}(1-\beta_\pm)u^2\right) \\
& \; + \frac{\nabla p_e}{ne} - \beta_\pm \frac{\nabla p}{ne} - \nabla \phi + \frac{\partial \nabla \psi_\pm}{\partial t} \; .
\end{align*}
Taking a cue from the Hall MHD results, we set $\phi_\pm = \phi - {\partial \psi_\pm}/{\partial t}$ and $\mathbf{E}_\pm = -\nabla \phi_\pm - (1/c)\partial \mathbf{A}_\pm / \partial t$.  We also note $\alpha_\pm \beta_\pm = \mu = m_e / m_i$, $1-\beta_\pm = -\alpha_\pm$, and $\beta_\pm = - \alpha_\mp$. Together, all these identities allow a considerable simplification:
\begin{equation} \label{OhmsLawPM}
\mathbf{E}_\pm + \frac{\mathbf{v}_\pm \times \mathbf{B}_\pm}{c} = \frac{m}{e}\nabla \left(\frac12 \beta_\pm v^2_\pm\right)
+ \frac{1}{ne}\nabla \left(\beta_\pm p_i - \beta_\mp p_e \right) \; .
\end{equation}
Thus, instead of the differing \eqref{OhmsLawU} and \eqref{MomentumEquationU}, or the greatly different \eqref{MomentumEquation} and \eqref{OhmsLaw}, we have two highly symmetric versions of Ohm's Law expressed using our beloved advected quantities.  Because we derived \eqref{OhmsLawPM} only by taking linear combinations and applying lots of vector identities, no information has been lost, and the pleasant equations \eqref{OhmsLawPM} are fully equivalent to the messy \eqref{MomentumEquation} and \eqref{OhmsLaw}.  Finally, in the $\mu \rightarrow 0$ limit, $\beta_+ \rightarrow 1$ and $\beta_- \rightarrow 0$, which explains why the $\nabla(v^2/2)$ term appears in the Hall MHD momentum expression \eqref{ExtraOhmsLaw} but not in its standard Ohm's Law \eqref{HallOhmsLaw}.

\subsection{Action}
\label{subsec:ExMHDAction}

We now turn to the problem of writing a fully Lagrangian action principle for Extended MHD.  As in Ref. \cite{keramidas14}, this is done by retaining terms up to first order in $\mu = m_e / m_i$ in the action, and taking $\mu \ll 1$ after variations.  The coordinate change expressed in \eqref{CoordinateChange} now becomes
\begin{align*}
\mathbf{\dot{q}}_i = \mathbf{\dot{q}} - \frac{m_e}{m_i + m_e} \mathbf{\dot{q}}_d \approx & \; \mathbf{\dot{q}} - \mu \, \mathbf{\dot{q}}_d 
 \\
\mathbf{\dot{q}}_e = \mathbf{\dot{q}} + \frac{m_i}{m_i + m_e} \mathbf{\dot{q}}_d \approx & \; \mathbf{\dot{q}} + (1-\mu) \, \mathbf{\dot{q}}_d
\nonumber
\end{align*}
Expanding the standard Lagrangian \eqref{ElectronIonLagrangian} and keeping terms up to first order in $\mu$ gives a new Lagrangian density
\begin{align}
\mathcal{L} = & \; \frac12 m n_0 \left[(1-\mu)\dot{q}^2 - 2 \mu \mathbf{\dot{q}}_d \cdot \mathbf{\dot{q}} + \mu (\dot{q}')^2 
+ 2 \mu \mathbf{\dot{q}}'_d \cdot \mathbf{\dot{q}}' + \mu (\dot{q}'_d)^2 \right] \label{ExtendedTangentAction} \\
& + \frac{e n_0}{c}\left[ \mathbf{\dot{q}}\cdot\mathbf{A} - \mu \mathbf{\dot{q}}_d \cdot \mathbf{A} 
- \mathbf{\dot{q}}' \cdot \mathbf{A}' - (1-\mu) \mathbf{\dot{q}}'_d \cdot \mathbf{A}' \right] \nonumber \\
&  - e n_0 \left[\phi(q,t) - \phi(q' + q'_d,t) \right] 
 - n_{0} \left[ U_i  \left( \frac{n_{0}}{\mathcal{J}}, s_{(0)i} \right) + U_e \left(\frac{n_0}{\mathcal{J}_f}, s_{(0)e} \right) \right] \nonumber \; .
\end{align}
Here $\mathbf{A} \equiv \mathbf{A}(q,t)$ and $\mathbf{A}' \equiv \mathbf{A}(q' + q'_d, t)$.

While there are many new terms in this Lagrangian, remarkably few make it to the actual equations of motion.  This is partly due to the cancellations that occur (as in Hall MHD) after setting $a' = q_f^{-1}(q(a,t),t)$, and partly due to the ordering $\mu \ll 1$ imposed after variations.  The $q$ variation gives
\begin{equation*}
m n_0 \mathbf{\ddot{q}} + \nabla p + \frac{n_0 e}{c} \mathbf{\dot{q}}_d \times \mathbf{B} = 0
\end{equation*}
and the $q_d$ variation gives
\begin{equation*}
m \mu n_0 \mathbf{\ddot{q}}_d + e n_0 \left(\mathbf{E} + \frac{\mathbf{\dot{q}} \times \mathbf{B}}{c} \right) +
\frac{e n_0}{c} \mathbf{\dot{q}}_d \times \mathbf{B} + \nabla p_e = 0 \; .
\end{equation*}
These are the correct equations, given that $\mathbf{\ddot{q}}$ and $\mathbf{\ddot{q}}_d$ will be complicated expressions when Eulerianized, particularly the latter.  See Sec IV.A of Ref. \cite{keramidas14} for a detailed explanation on how to convert them into Eulerian form.

The Lagrangian density \eqref{ExtendedTangentAction} produces canonical momenta
\begin{equation*}
\pi^i = m n_0 \dot{q}^i \qquad \qquad \pi_d^i = \mu  m n_o \dot{q}^i_d - \frac{e n_0}{c} A^i \; .
\end{equation*}
Thus
\begin{equation} \label{BPhaseSpace}
\mathbf{B}_\star = - \frac{c}{n_0 e} \nabla \times \mathbf{\pi}_d \qquad \qquad
\mathbf{B}_\pm = \frac{c}{n_0 e} \left( - \nabla \times \mathbf{\pi}_d + \beta_\pm \nabla \times \mathbf{\pi} \right) \; .
\end{equation}
The $\mathbf{B}_\pm$ are advected by $\mathbf{v}_\pm$, whose Lagrangian equivalents we will call $\mathbf{\dot{q}}_\pm \equiv \dot{\mathbf{q}} - \alpha_\pm \dot{\mathbf{q}}_d $, where the minus sign comes because $\mathbf{u}$ and $\dot{\mathbf{q}_d}$ differ by a sign (see \eqref{PiDVariation}).  Thus we have these fully expanded expressions for $\mathbf{B}_\pm$, in analogy with \eqref{FluxConservation}:
\begin{equation} \label{BPMAdvection}
\mathbf{B}^i_\pm = \frac{\mathbf{B}^j _{(0)\pm}}{\mathcal{J}_\pm} \frac{\partial q^i_\pm}{\partial a^j}
= \frac{\mathbf{B}^j _{(0)\pm}}{\mathcal{J}_\pm} \left(\frac{\partial q^i}{\partial a^j} - \alpha_\pm \frac{\partial q^i_d}{\partial a^j}\right) \; ,
\end{equation}
where $q_\pm(a',t)$ are the integral lines of $\dot{q}_\pm$, and $\mathcal{J}_\pm$ are the determinants of the matrices $\partial q_\pm / \partial a$.

Producing a phase-space action will involve bringing in the omitted term $\int (B^2 / 8 \pi) d^3 x$ (see the note about this term in the Hall MHD section), and converting the terms $\mathbf{\dot{q}}_d\cdot\mathbf{A}$ and $\dot{q}_d^2 /2$ to field terms by anticipating the relation $(4\pi/c) \mathbf{j} = \nabla \times \mathbf{B}$.  By integrating by parts on the $\dot{q}^2_d / 2$ term, one can combine both of them into the single term $(1/8\pi)\mathbf{B}\cdot \mathbf{B}_\star$.  But first we need to show how to actually perform a variation on such a term.

In Eulerian variables we have
\begin{equation*}
\mathbf{B}_\star = \mathbf{B} + \frac{m_e c}{e} \nabla \times \left(\frac{\mathbf{j}}{ne} \right)
= \mathbf{B} + \frac{m_e c}{e^2} \nabla \times \left(\frac{\nabla \times \mathbf{B}}{n} \right) \; ,
\end{equation*}
so that $\mathbf{B}_\star$ is a function of $\mathbf{B}$ and $n$.  Thankfully, the situation is simpler in Lagrangian variables.  The variable appears inside a $d^3 q$ integral, contracted with another vector.  One can integrate by parts to remove a curl, swap in $d^3 q = \mathcal{J}_f d^3 a$ (any term involving $\mathbf{B}_\star$ will be an electron quantity), and integrate by parts again, so that
\begin{equation*}
\mathbf{B}_\star \rightarrow \mathcal{J}_f \mathbf{B} + \frac{m_e c}{n_0 e^2} \nabla \times \left(\nabla \times \mathbf{B}\right)
\end{equation*}
whenever it appears inside a label space integral.  Thus we have $\mathbf{B}_\star (\mathbf{B})$ in Lagrangian variables.  We assume the differential equation can be inverted to produce
\begin{equation} \label{BBStar}
\mathbf{B}^i(q) = \int \mathbf{B}^j_\star(q')\;  G^{ij} (q; q') \; d^3 q'
\end{equation}
for some tensorial Green's function $G^{ij}$.  Therefore, when varying $(1/8\pi)\mathbf{B}\cdot\mathbf{B}_\star$, the result is
\begin{align}
\int \delta \left(\frac{\mathbf{B}\cdot\mathbf{B}_\star}{8\pi}\right) \; \mathrm{d}^3 q = & \;
\frac1{8\pi} \int \! \int \delta \left( B^i_\star (q) B^j_\star (q') G^{ij}(q;q') \right) \mathrm{d}^3 q' \; \mathrm{d}^3 q \nonumber \\
= & \; \frac1{8\pi} \int \! \int \left( \delta B^i_\star (q) B^j_\star (q') G^{ij}(q;q') + B^i_\star (q) \delta B^j_\star (q') G^{ij}(q;q') \right) 
\mathrm{d}^3 q' \; \mathrm{d}^3 q \nonumber \\
= & \; \frac{1}{4\pi} \int \! \int \left( \delta B^i_\star (q) B^j_\star (q') G^{ij}(q;q') \right) \mathrm{d}^3 q' \; \mathrm{d}^3 q 
= \frac{1}{4\pi} \int \mathbf{B} \cdot \delta \mathbf{B}_\star \; \mathrm{d}^3 q
\label{VariationBBStar}
\end{align}
where we have used the symmetry $G^{ij}(q;q') = G^{ij}(q';q)$ of Green's functions. The variation did not affect $G^{ij}(q;q')$ despite the dependence on $q$ because the Green's function is translation invariant, i.e. $G^{ij}(q; q') = G^{ij}(\mathbf{q}-\mathbf{q}')$.

We are now in a position to write the full phase-space Lagrangian.  It is
\begin{align}
L = & \; \int \! \int \left[\pi^i \dot{q}^i + \pi^i_d \dot{q}^i_d - \frac{\pi^2}{2 m n_0} + 
\frac{e}{mc} \left(\pi^i A^i(q,t) - \pi^i  A^i(q' + q'_d,t) \right) - e n_0 \Big(\phi(q,t) \right. \nonumber \\
& \; \quad \left.
 - \phi(q' + q'_d,t)\Big) - n_0 \left(U_i \left(\frac{n_0}{\mathcal{J}},s_{(0)i}\right) + U_e \left(\frac{n_0}{\mathcal{J}_f},s_{(0)e}\right) \right)
\right] \delta(a' - q_f^{-1}(q(a,t),t)) \nonumber \\
& + \; \frac{J_f}{8\pi (\Delta \beta)^2} \left[ \frac{\beta_-}{\mathcal{J}_+}
\left(\nabla \times \pi_d - \beta_+ \nabla \times \pi \right)^k 
\left(\frac{\partial q^i}{\partial a^k} - \alpha_+ \frac{\partial q^i_d}{\partial a^k}\right) \right. \nonumber \\
&\; \left. \quad - \frac{\beta_+}{\mathcal{J}_-} \left(\nabla \times \pi_d - \beta_- \nabla \times \pi \right)^k 
\left(\frac{\partial q^i}{\partial a^k} - \alpha_- \frac{\partial q^i_d}{\partial a^k}\right)
\right] \label{ExtendedPhaseSpaceAction} \\
& \; \times \left[ \frac{\beta_-}{\mathcal{J}_+}
\left(\nabla \times \pi_d - \beta_+ \nabla \times \pi \right)^l 
\left(\frac{\partial q^j}{\partial a^l} - \alpha_+ \frac{\partial q^j_d}{\partial a^l}\right) \right. \nonumber \\
& \; \left. \quad - \frac{\beta_+}{\mathcal{J}_-} \left(\nabla \times \pi_d - \beta_- \nabla \times \pi \right)^l 
\left(\frac{\partial q^j}{\partial a^l} - \alpha_- \frac{\partial q^j_d}{\partial a^l}\right)
\right] G^{ij}(q;q') \; \mathrm{d}^3 a \; \mathrm{d}^3 a' \; , \nonumber
\end{align}
where we have set $\Delta \beta \equiv \beta_+ - \beta_-$ and expanded $(1/8\pi)\mathbf{B}\cdot\mathbf{B}_\star$ using \eqref{BBStar}, the first equation of \eqref{BStarFromBPM}, and \eqref{BPMAdvection}.  
Thankfully, variations are simplified considerably by the result \eqref{VariationBBStar}.  It is also interesting that the usual delta function is replaced by a more specialized Green's function in the $(1/8\pi)\mathbf{B}\cdot\mathbf{B}_\star$ term.  It is worth pointing out that, while we assumed barotropic equations of state in our development, the above action works for the more general equations of state $U_s(n,s)$.

The $\pi$ variation gives
\begin{equation} \label{ExPiVariation}
\dot{q}^i = \frac{\pi^i}{m n_0} \; .
\end{equation}
The $\pi$ variations occurring in the big field term all cancel, which is not surprising since $\mathbf{B}_\star$ has no $\pi$ dependence.  The $\pi_d$ variation gives
\begin{equation} \label{ExPiDVariation}
\mathbf{\dot{q}}_d = - \frac{c}{4 \pi n_0 e} \nabla \times \mathbf{B}
\end{equation}
as it should, with the factor of $\mathcal{J}_f$ being absorbed back into $\mathrm{d}^3 q$ when invoking \eqref{VariationBBStar}.  After some work, the $q$ variation gives
\begin{equation*}
- \mathbf{\dot{\pi}}^i - \frac{C^{kj}}{8 \pi} \frac{\partial}{\partial a^k} \left[B^i_\star B^j + B^i B^j_\star - B_\star^k B^k \delta^{ij} \right] 
+ \mathcal{J} \nabla^i p = 0 \; .
\end{equation*}
Imposing the $\mu \ll 1$ condition turns this into
\begin{equation} \label{ExQVariation}
- \mathbf{\dot{\pi}}^i - \frac{C^{kj}}{4\pi} \frac{\partial}{\partial a^k} \left[B^i B^j - \frac{B^2}{2} \delta^{ij} \right] + \mathcal{J} \nabla^i p = 0 \; .
\end{equation}

Finally, the $q_d$ variation gives
\begin{align*}
0 = & \;  - \dot{\pi}_d^i - \frac{e}{mc} \pi^j \partial^i A^j +n_0 e\partial^i \phi + \mathcal{J}_f \partial^i p_e \\
& \; - \frac{C^{kj}}{8\pi \Delta\beta} \frac{\partial}{\partial a^k} 
\left[B^i (\alpha_+ \beta_- B_+ - \alpha_- \beta_+ B_-)^j + B^j (\alpha_+ \beta_- B_+ - \alpha_- \beta_+ B_-)^i \right. \\
& \; \left. - B^m (\alpha_+ \beta_- B_+ - \alpha_- \beta_+ B_-)^m \delta^{ij} \right] \; .
\end{align*}
To simplify this, note that
\begin{equation*}
\alpha_+ \beta_- \mathbf{B}_+ - \alpha_- \beta_+ \mathbf{B}_- = \left(\alpha_+ \beta_-  - \alpha_- \beta_+ \right) \mathbf{B}_\star
+ \frac{mc}{e} \left(\alpha_+ - \alpha_- \right)\beta_- \beta_+ \left(\nabla \times \mathbf{\dot{q}}\right) \; .
\end{equation*}
Now $\beta_- \beta_+ = - \mu$, so we end up dropping the $\nabla \times \mathbf{\dot{q}}$ term.  Next $\alpha_+ \beta_-  - \alpha_- \beta_+ = -\sqrt{1+4\mu} \approx -1$, so we get a plain $-\mathbf{B}_\star$ term, which again gets reduced to $-\mathbf{B}$.  In the end,
\begin{equation} \label{ExQDVariation}
- \dot{\pi}_d^i - \frac{e}{mc} \pi^j \partial^i A^j + \frac{C^{kj}}{4\pi} \frac{\partial}{\partial a^k} \left[B^i B^j - \frac{B^2}{2} \delta^{ij} \right]
+ n_0 e\partial^i \phi + \mathcal{J}_f \partial^i p_e = 0
\end{equation}
Remarkably, equations \eqref{ExPiVariation}, \eqref{ExPiDVariation}, \eqref{ExQVariation}, and \eqref{ExQDVariation} are identical to their equivalents \eqref{PiVariation}, \eqref{PiDVariation}, \eqref{QVariation} and \eqref{QDVariation} from Hall MHD.  This is likely the source of the maps discovered in Ref. \cite{lingam15}.  The differences between Hall and Extended MHD arise when you switch to Eulerian variables.

\subsection{Derivation of the bracket}
\label{subsec:ExMHDELMap}

The field variable $\mathbf{B}^*$ will be written as a linear combination of the two-forms $\mathbf{B}_{\pm}$, each of which is advected along a linear combination of $q$ and $q_d$.  Thus \eqref{MagneticEulerLagrange} will be rewritten as
\begin{align} \nonumber
B^i_\star (x,t) = & \int \left(-\frac{\beta_-}{\Delta \beta}\right) B_{(0)+}^{j} (a) \left(\frac{\partial q^i}{\partial a^j} - \alpha_+ \frac{\partial q_d^i}{\partial a^j} \right) 
\delta(x - q(a,t) + \alpha_+ q_d(a,t)) \: \\
& + \frac{\beta_+}{\Delta \beta} B_{(0)-}^{j} (a) \left(\frac{\partial q^i}{\partial a^j} - \alpha_- \frac{\partial q_d^i}{\partial a^j} \right) 
\delta(x - q(a,t) + \alpha_- q_d(a,t)) \: d^3 a \; .
\label{BStarExpanded}
\end{align}
The coefficients $\alpha_\pm$ are given by \eqref{Coefficients}, and those in front of the $B_{(0)\pm}$ come from inverting \eqref{LinearCombination}, writing $\Delta \beta = \beta_+ - \beta_-$.  The minus signs in front of the various $\alpha_\pm$ arise because $\mathbf{\dot{q}}_d$ and $\mathbf{u}$ (via \eqref{ExPiDVariation}) differ by a sign.  

Equation \eqref{BPhaseSpace} shows that
\begin{equation*}
\delta \mathbf{B}_{(0)\pm} = \frac{c}{n_0 e} \left(-\nabla \times \delta \pi_{(0)d} + \beta_\pm \nabla \times \delta{\pi}_{(0)}\right) \; .
\end{equation*}
However, despite the appearance of $\delta \pi_{(0)}$ here, the expression $\delta f / \delta \pi$ is unchanged, because the terms arising from the two parts of $\delta \mathbf{B}_\star$ cancel each other.  However, $\delta f / \delta \pi_d$ is changed, converting $\{f,g\}_{Hall}$ into its Extended MHD counterpart.

The following changes appear in the previous calculation: (i) All functional derivatives with respect to $\mathbf{B}$ are now done with respect to $\mathbf{B}_\star$; (ii) in the magnetic portion of $\{f,g\}_{MHD}$, $\mathbf{B}$ is replaced by $\mathbf{B}_\star$; (iii) in $\{f,g\}_{Hall}$, $\mathbf{B}$ is replaced by $\left[c/(n_0 e \Delta \beta)\right] ( \beta_- \alpha_+ \mathbf{B}_+ - \beta_+ \alpha_- \mathbf{B}_-)$.  This quantity works out to be $(c / n_0 e) (\mathbf{B}_\star - (m_e c / e) \nabla \times \mathbf{\dot{q}})$.  Thus we derive the following bracket:
\begin{equation*}
\{f,g\}_{ExMHD} = \{f,g\}_{MHD} + \{f,g\}_{Ex}
\end{equation*}
with $\{f,g\}_{MHD}$ given by the first part of \eqref{BigBracket} and
\begin{equation*}
\{f,g\}_{Ex} = \int - \frac{c}{n e} \left(\mathbf{B}_\star - \frac{m_e c}{e} \nabla \times \left(\frac{\mathbf{m}}{\rho }\right)  \right) 
\cdot \left[\left(\nabla \times \frac{\delta f}{\delta \mathbf{B}_\star}\right) \times
\left(\nabla \times \frac{\delta g}{\delta \mathbf{B}_\star}\right)\right] \: d^3x,
\end{equation*}
matching \eqref{XMHDBracket}.  As expected, the limit $\mu = (m_e/m_i) \rightarrow 0$ reduces $\mathbf{B}_\star$ to $\mathbf{B}$, eliminates the $\nabla \times \mathbf{v}$ term, and thus reduces $\{f,g\}_{Ex}$ to $\{f,g\}_{Hall}$.

\section{Conclusion}
\label{sec:Conclusion}

We have accomplished many things in the course of deriving the noncanonical brackets for Hall and Extended MHD.  The need for canonical momenta to serve as the backbone of the brackets led to actions for both theories: first the tangent-space ones \eqref{TangentSpaceAction} and \eqref{ExtendedTangentAction}, then the phase-space ones \eqref{PhaseSpaceDensity} and \eqref{ExtendedPhaseSpaceAction}.  Essential to the actions were the advected generalized vorticities: the magnetic field and \eqref{HallIonVorticity} for Hall MHD, and the two expressions \eqref{LinearCombination} for Extended MHD.  To go with these advected two-forms, we found advected one-form potentials \eqref{HallIonPotential} and \eqref{ExtendedPotentials}.  For Hall MHD, we found that the momentum equation could be restated in the form of an additional equation \eqref{ExtraOhmsLaw} resembling Ohm's Law.  Similarly, in Extended MHD, after recasting the complex original equations \eqref{MomentumEquation} and \eqref{OhmsLaw} into variables based around the various advected forms, we could produce the equivalent, but much simpler expressions \eqref{OhmsLawPM}.  Knowing these forms, we were also able to define the natural Euler-Lagrange maps \eqref{MagneticEulerLagrange} and \eqref{BStarExpanded}, and at last derive the noncanonical brackets \eqref{HallBracket} and \eqref{XMHDBracket}.

A number of interesting concepts had to be used in order to produce these results.  To begin with, the actions required a double label space in order to be fully Lagrangian, and we may speculate that Lagrangian theories incorporating $n$-fluid effects will require $n$ label spaces.  Moreover, Padhye and Morrison \cite{PMor96a,PMor96b} showed that, in ideal MHD, magnetic helicity is the Noether invariant corresponding to the symmetry of relabelling.  In each of Hall and Extended MHD, we have two label spaces, and we speculate that helicities corresponding to each theory's two generalized vorticities will arise from distinct relabelling symmetries on the doubled label space.  This is a matter for future research.  Next, our expanded inventory of two-forms and one-forms, defined by their advection properties, allowed us to greatly simplify Extended MHD.  This shows that a firm understanding of the geometric nature of the objects appearing in a physical system will allow one to cut away much of its seeming complexity, as briefly noted in \cite{lingam15}.  

Finally, our implementation of the Euler-Lagrange map (and the phase-space action principle) required an unusual method of implementing a constraint, a method which may turn out to have broader applicability. Prior to attempting work such as ours, one might have objected that Hall and Extended MHD are theories too specialized and inelegant to be a fruitful topic for mathematical physics. Fortunately, we have found that it is precisely when investigating such specialized problems that one may discover ideas and methods useful in a broader context.

From a practical perspective, we emphasize that our action principle for Extended MHD (and the concomitant noncanonical Hamiltonian formulation) can be applied to many problems of interest and relevance in diverse areas. We list a few examples in this category: topological invariants \cite{LMM16}, particle relabelling symmetries \cite{PMor96a,PMor96b,Ara15,Ara16}, reconnection based on Hamiltonian models \cite{HMIYA13,CGWB13,HHM15}, tearing modes \cite{YD12}, Hamiltonian closures \cite{PCPT14,PCPT15}, nonlinear waves \cite{Yo12,AY16}, weakly nonlinear dynamics \cite{Hir11,MV16}, the derivation of gyrofluid and hybrid fluid-kinetic models \cite{Tr10,MLA14,LMor14,TTCM14,KWM15}, the properties of the equatorial electrojet \cite{HHMH16}, and the rapidly burgeoning field of variational integrators \cite{Het12,NVB13,ZQBY14}.

Thus, our work serves to advance and flesh out the mathematical foundations of Extended MHD, whilst also paving the way for the applications of our methodology in fusion, space and astrophysical plasmas. 

\section*{Acknowledgments}
PJM and ECD were supported by DOE Office of Fusion Energy Sciences, under DE-FG02-04ER-54742. ML was supported by the NSF Grant No. AGS-1338944 and the DOE Grant No. DE-AC02-09CH-11466.


%

\end{document}